\documentclass[ ]{iopart}
\usepackage{graphicx}
\usepackage{iopams}

\begin{document}
\title{Formation and interaction of resonance chains in the open 3-disk system}

\author{T.~Weich$^{1,2}$, S.~Barkhofen$^2$, U.~Kuhl$^{3,2}$, C.~Poli$^4$, and H.~Schomerus$^4$}
\address{$^1$ Fachbereich Mathematik, Philipps-Universit\"{a}t Marburg, Hans-Meerwein-Stra{\ss}e,
35032 Marburg, Germany}
\address{$^2$Fachbereich Physik, Philipps-Universit\"{a}t Marburg, Renthof 5,
35032 Marburg, Germany}
\address{$^3$Laboratoire de Physique de la Mati\`{e}re Condens\'{e}e, CNRS UMR 7336, Universit\'{e} de Nice
Sophia-Antipolis, F-06108 Nice, France}
\address{$^4$Department of Physics, Lancaster University, Lancaster LA1 4YB, United Kingdom}
\ead{ulrich.kuhl@unice.fr}

\date{\today}
\begin{abstract}
In ballistic open quantum systems one often observes that the resonances in the complex-energy plane form a clear chain structure. Taking the open 3-disk system as a paradigmatic model system, we investigate how this chain structure is reflected in the resonance states and how it is connected to the underlying classical dynamics. Using an efficient scattering approach we observe that resonance states along one chain are clearly correlated while resonance states of different chains show an anticorrelation. Studying the phase space representations of the resonance states we find that their localization in phase space oscillate between different regions of the classical trapped set as one moves along the chains and that these oscillations are connected to a modulation of the resonance spacing. A single resonance chain is thus no WKB quantization of a single periodic orbits, but the structure of several oscillating chains arises from the interaction of several periodic orbits. We illuminate the physical mechanism behind these findings by combining the semiclassical cycle expansion with a quantum graph model.
\end{abstract}


\maketitle

\section{Introduction}
\label{sec:Introduction}

The study of geometrically open (leaky) quantum systems is a subject of intense research, with recent interest driven in equal parts by concrete applications (including electronic transport \cite{bee97,naz09} and microcavity lasers \cite{gma98,har11}) and deep fundamental questions \cite{alt13,nov13,non11}. Many of the systems of interest display the signatures of quantum chaos, which in the presence of leakage become enriched by the formation of a fractal trapped set in classical limit. Over the past decade, it has been realised that this phenomenon finds a quantum-mechanical analogue in the distribution of the resonances in the complex-energy plane. The localization of resonance states on the classical trapped set was already observed by Casati, Maspero and Shepelyanski \cite{cas99a}, who numerically studied the open Chirikov map. Mathematical work established that in consequence the resonances follow a modified, fractal, Weyl law \cite{sjo90,zwo99b,lu03,gui04,non05,sjo07}. This fractal Weyl law has been confirmed numerically for a wide range of quantum maps \cite{sch04e,str04,chr07,ram09,ped09,kop10,erm10,erm11,ped12,kor13}, and numerical \cite{wie08a,sch09,ebe10,arXbor13} as well as first experimental \cite{pot12} work shows that the fractal Weyl law also holds for autonomous systems. In a simple physical picture, this law originates from the increasing phase space resolution as one approaches the classical limit, which results in a proliferation of short-living resonances that follow ballistical classical escape routes \cite{sch04e}. The observations of Casati, Maspero and Shepilyanski on the localization of resonance states have been refined by Keating \etal \cite{kea06b} and Nonnenmacher and Rubin \cite{non07}, who distinguished between the left and right resonance states, and showed for the baker map that all semiclassical measures (i.e., all possible semiclassical limits of these states) localize on the forward respectively the backward trapped set. Later this localization was rigorously shown by
Nonnenmacher and Zworski \cite{non09} for a much larger class of systems. A classification of the possible semiclassical measures similar to the Schnirelman theorem for closed systems is still open and has so far only been obtained for the very special case of the Walsh quantized baker map \cite{non07, kea08b}.

Autonomous systems are much richer than maps as they display an additional, even more ubiquitous phenomenon related to quantum-to-classical correspondence, namely, the formation of resonance chains encountered in various systems such as quantum cavities with large openings \cite{naz02,bul06}, dielectric microresonators with low index materials \cite{wie03,leb06,leb08,wie08,ryu13}, and models of mathematical interest such as convex co-compact surfaces (Schottky surfaces) \cite{arXbor13}). The simplest resonance chains are supported by a single ballistic trajectory, resulting in a sequence of states in close analogy to those of a particle in a one-dimensional box with leaking walls \cite{ber72}. However, for open chaotic systems the physical picture behind the fractal Weyl law implies that the states on such chains must eventually morph into complex wave-chaotic states supported by the trapped set. In this work we make a first step towards an understanding of the question how these fundamental phenomena fit together. Firstly, we show that the clearly distinguishable resonance chains display systematic interactions driven by the quantization of more than just a single trajectory. These interactions manifest themselves in systematic oscillations of resonance spacings and life times, which go along with crossings of the chains in the complex plane and oscillations of the phase-space support of the associated resonance states. Secondly, we show that the initially well separated pairs of resonance chains merge and perform more complicated interactions as one moves further into the semiclassical limit. We also observe that the first merging of the chains is linked to a better phase space resolution of the trapped set.

We develop this picture by considering a paradigmatic open autonomous system with a fractal trapped set, the 3-disk system, which has been introduced by Gaspard and Rice \cite{gas89b,gas89a,gas89d} and Cvitanovic and Eckhardt \cite{cvi89}, and has found several experimental realizations \cite{lu99,pan00a,pot12,bar13b}. We employ an efficient scattering approach that allows us to calculate resonance states. Via these resonance states and the corresponding phase
space representations we can study their localization on classical phase space structures and the connection to the resonance chain structure. This leads us to the observation that resonances on one chain show a significantly different correlation behavior than resonances on different chains. The chain structure is thus not only a visual impression but reflects common physical properties. In order to extract the underlying mechanism behind these correlations and modulations, we formulate an approximate correspondence of the 3-disk system with an open quantum graph, consisting of two edges of different lengths. This correspondence holds in leading order of the cycle expansion, and implies that the systematic correlations and interactions of the chains originate from the approximate commensurability of the fundamental periodic orbits in the system. Deeper in the semiclassical limit, we find that the merging of resonance chains coincides with an increased population of the classical trapped set by long-living resonance states.

This paper is organised as follows. In \sref{sec:model} we review the key classical and quantum features of the 3-disk system and describe our method to calculate the full spatial structure of the resonance states. Furthermore, we explain how this method can be used to efficiently calculate Poincar\'{e}-Husimi distributions, as well as the symmetric phase space distributions proposed by Ermann \etal \cite{erm09}. These methods are then used in \sref{sec:results} to relate the resonance states to the underlying classical phase space structures, allowing us to extract the spatial correlations that complement the modulations of the spectral features. In \sref{sec:quantumGraph} we describe the approximate correspondence of the 3-disk system with the open quantum graph and discuss how this leads to a phenomenological explanation of our numerical observations. \Sref{sec:conclusions} contains our conclusions.

\section{Model and methods}
\label{sec:model}

\subsection{The 3-disk system}

\begin{figure}
  \centering
  \includegraphics[width=0.6\columnwidth]{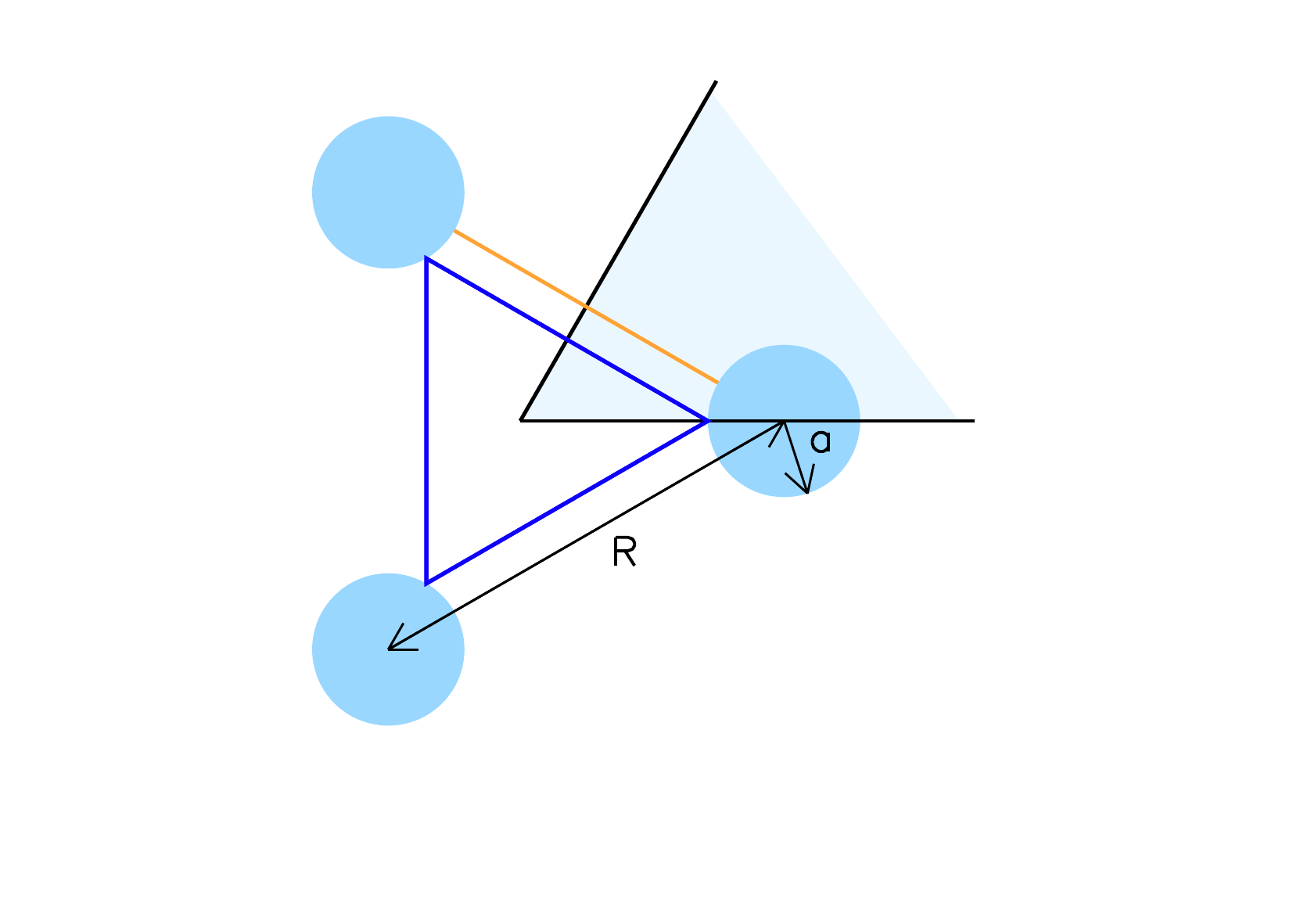}\\
  \hspace*{1cm}\includegraphics[width=0.9\textwidth]{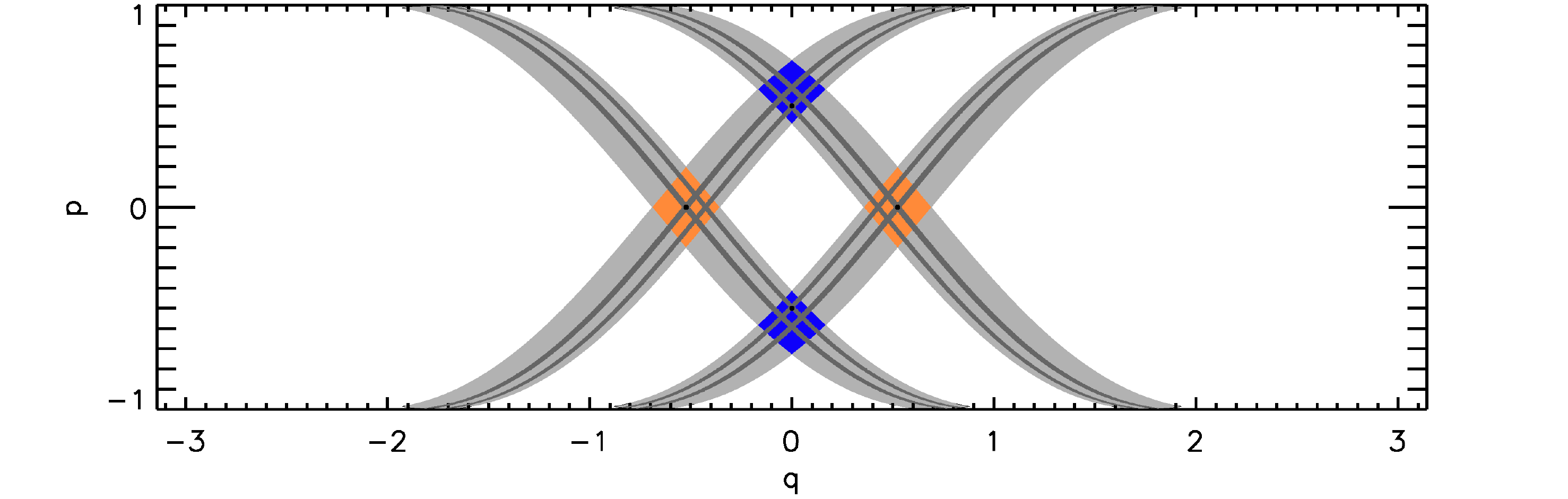}\\
  \includegraphics[width=0.9\textwidth]{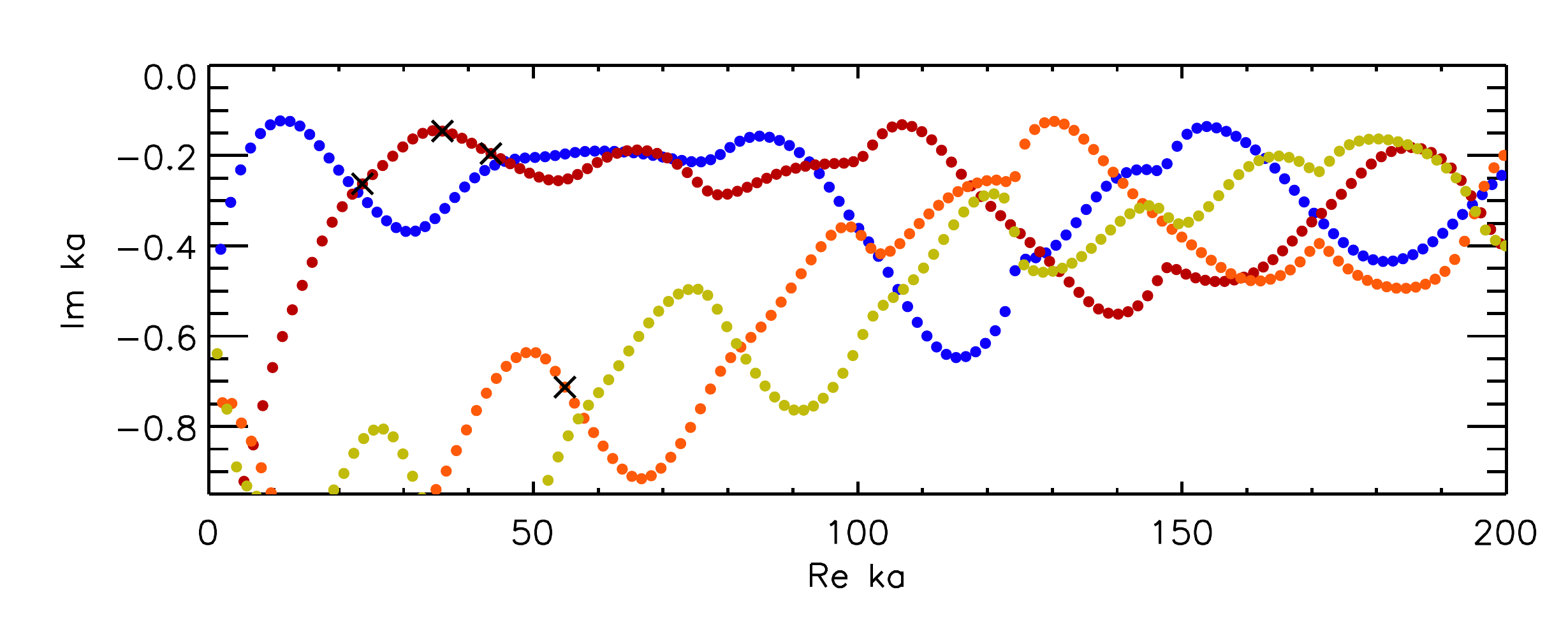}
\caption{\label{fig:3Disk_trappedSet}\label{fig:3Disk_Orbits}\label{fig:Res_chains}
Top panel: Schematic drawing of the 3-disk system (with the fundamental domain shaded in light blue), consisting of three disks of radius $a$ centered at the corners of an equilateral triangle of side length $R$, where here and throughout this work $R/a$ = 6. Also shown are the two fundamental orbits $0$ (orange) and $1$ (blue). Middle panel: forward and backward trapped sets of order $n=1$ in light grey and $n=2$ in dark grey, displayed in the Birkhoff coordinates of one of the three disks. The intersections of these sets make up the trapped sets $\mathcal T^{(n)}$. The regions $\mathcal T^{(1)}_i$ visited by the orbits $i=0,1$ are shaded in orange and blue, respectively; the location of these orbits [$(q,p)_{ 0} = (\pm 1/6\pi, 0)$ and $(q,p)_{1} = (0, \pm 0.5)$] is indicated by the black dots. Bottom panel: Position of the quantum-mechanical resonances $k_n$ of the $A_2$-symmetry reduced system in the complex-$k$ plane. The resonances organize into chains, which are marked in different colors. The crosses mark some resonances examined in more detail later on in this work.}
\end{figure}

In the 3-disk system \cite{gas89b,gas89a,gas89d,wir99b} a particle moves in a two-dimensional plane and scatters off identical hard disks whose centers form an equilateral triangle (see \fref{fig:3Disk_Orbits}). The distance between the centers of the disks is denoted by $R$, and the disk radius is denoted by $a$. Up to scaling the geometry is completely defined by the ratio $R/a$.

The quantum system is described by the free Schr\"{o}dinger equation
\begin{equation}\label{eq:Schr_GL}
-\nabla^2\psi(\mathbf{r})=k^2\psi(\mathbf{r})
\end{equation}
with Dirichlet boundary conditions at the disk boundaries. Further imposing outgoing boundary conditions on the scattering states, the spectrum becomes a discrete set of quantum resonances with complex wave number $k_n=\mathrm{Re}\,k_n+\rmi\mathrm{Im}\,k_n$ and associated resonance wave functions $\psi_n({\bf r})$. The complexness of the resonances $k_n$ arises from the non-hermiticity of the wave equation with outgoing boundary conditions, and thus reflects the leakage out of the system.

The disk configuration of the 3-disk system is symmetric under the symmetry group $C_{3v}$ and allows a symmetry reduction which is well understood on the classical and semiclassical level \cite{cvi93} as well as on the quantum level \cite{gas89d}. We focus on resonance states in the $A_2$-representation, which is the reduction of choice for experimental realizations of the desymmetrized system. These states have to be antisymmetric with respect to reflections at the symmetry axes, and thus describe the resonances in the fundamental domain (shaded in light blue in \fref{fig:3Disk_Orbits}) with additional Dirichlet boundary conditions at the boundaries of the fundamental domain. The resonances can be calculated numerically from the poles of the scattering matrix, which can be expressed
in terms of Bessel and Hankel functions \cite{gas89d} (see also \sref{sec:ScatStates}). As shown in the bottom panel of \fref{fig:Res_chains}, these resonances form well-defined, systematically interacting chains, whose appearance is an ubiquitous feature in open quantum and wave-optical systems.

We aim to explain the formation and interaction of these chains in terms of the underlying classical phase space structures. Classically, the particle moves ballistically along straight lines, with specular reflections off the hard walls of the disks. The escape out of the system is then conveniently analysed via the trapped sets--the forward trapped set of order $n$, which contains all points in phase space which have at least $n$ reflections on the disks in positive time direction, and the backward trapped set of order $n$, which is defined analogously for the negative time direction. The intersection of the forward and backward trapped set makes up the trapped set of order $n$, denoted by $\mathcal T^{(n)}$. In the limit $n\to \infty$, these sets form Cantor sets (the forward trapped set, the backward set, and the trapped set, respectively).

In the 3-disk system, the trapped sets are conveniently examined on the Poincar\'{e} section of boundary reflections at one of the disks, parametrized by Birkhoff coordinates $q \in [-\pi,\pi]$ (the arclength around the disk, expressed in units of $a$) and $p=\hat \mathbf{t} \cdot \hat \mathbf{p} \in [-1,1]$ (the projection of the normalised momentum $\hat \mathbf{p}$ onto the tangent $\hat \mathbf{t}$). The trapped sets of order 1 and 2 are shown in the middle panel of \fref{fig:3Disk_Orbits}. The backbone of these sets is formed by the periodic orbits in the system. It is known that for sufficiently separated disks the set of periodic orbits can be described by a complete symbolic dynamic of the alphabet $\{0,1\}$ \cite{cvi89}. The two fundamental orbits are given by the two words of length one, where $0$ corresponds to the bouncing ball orbit, and $1$ corresponds to the triangular orbit (see \fref{fig:3Disk_Orbits}). An analogous symbolic dynamics applies to the trapped sets \cite[Section III]{gas89b}. We will only need to distinguish the four regions of trapped set $\mathcal T^{(1)}$, and denote the union of the two regions around the orbit $0$ with $\mathcal T^{(1)}_0$ and the two regions around orbit $1$ with $\mathcal T^{(1)}_1$.

\subsection{Calculation of scattering states}
\label{sec:ScatStates}

We obtain the resonance states $\psi_n$ by extending the approach by Gaspard and Rice, who determined the resonance wave numbers $k_n$ from the poles of the scattering matrix $S$ \cite{gas89d,wir99b}. Far from the scattering center, the scattering state takes the asymptotic form
\begin{equation}\label{eq:psi_asym}
\psi_{l} (\mathbf{r};k) \approx ~\sum_{l^\prime = -\infty}^{\infty}\left[~\delta_{ll^\prime}~\psi^{\mathrm{in}}_{l^\prime}(\mathbf{r};k)
+S_{ll^\prime}(k)~\psi^{\mathrm{out}}_{l^\prime}(\mathbf{r};k) \right]
\end{equation}
where
\[
 \psi^{\mathrm{in}}_{l^\prime} (\mathbf{r};k):= \frac{1}{\sqrt{2\pi kr}}\exp(-\rmi(kr-l^\prime\pi/2-\pi/4))\exp(\rmi l^\prime\phi)
\]
and
\[
\psi^{\mathrm{out}}_{l^\prime}(\mathbf{r};k) := \frac{1}{\sqrt{2\pi kr}}\exp(\rmi(kr-l^\prime\pi/2-\pi/4))\exp(\rmi l^\prime\phi)
\]
are the asymptotic forms of incoming and outgoing waves with angular momentum $l^\prime$, while the scattering matrix $S_{ll^\prime}(k)$ describes the coupling between these states. The scattering matrix extends meromorphically to the complex-$k$ plane, and the resonances are precisely given by its poles.

Let now $k_n \in \mathbb{C}$ be a resonance of multiplicity one and $k\in \mathbb C$ a nearby point. After truncating the $S$-matrix for sufficiently high angular momenta, $S_{ll^\prime}(k)$ becomes a well defined matrix and there is a maximal singular value $\sigma(k)$ as well as two normalized vectors $\vec i(k)=(i_{l}(k))_{l \in \mathbb Z}$ and $\vec a(k)=(a_{l}(k))_{l\in \mathbb Z}$ such that
\[
\sum_{l} i_{l}(k)~S_{ll^\prime}(k)=\sigma(k) a_{l^\prime}(k).
\]
As $S_{ll^\prime}(k_n)$ is singular, the maximal singular value $\sigma(k)$ will diverge for $k\to k_n$. In order to define the resonant scattering state we assume that in this limit, the two vectors $\vec i(k)$ and $\vec a(k)$ converge to two well defined vectors $\vec i$ and $\vec a$. Symbolically we write
\begin{equation}\label{eq:singular_S}
\sum_{l} i_{l}~S_{ll^\prime}(k_n) =\infty~ a_{l^\prime} =\alpha_{l^\prime}.
\end{equation}
In practice we will never be able to evaluate the scattering matrix exactly at the pole, so also $\vec i$ and $\vec a$ will be approximations to the idealized singular vectors, and \eref{eq:singular_S} practically means that the corresponding singular value is so large that all terms that do not contain $\alpha_{l^\prime}$ can be neglected \cite{tre97}.

With this singular vector $\vec i$ we define the asymptotic behavior of the resonant scattering state to the pole $k_n$ as
\begin{eqnarray} \label{eq:psi_scat}
\psi_n (\mathbf{r}) &:=& \sum_{l} i_{l} ~\psi_{l}(\mathbf{r};k_n) \\ \nonumber
			&=& \sum_{l}~i_{l}~\sum_{l^\prime = -\infty}^{\infty}\left[~\delta_{ll^\prime}~\psi^{\mathrm{in}}_{l^\prime}(\mathbf{r};k_n)
+S_{ll^\prime}(k_n)~\psi^{\mathrm{out}}_{l^\prime}(\mathbf{r};k_n) \right]\\ \nonumber
			&=& ~\left[ \sum_{l}~i_{l} ~\psi^{\mathrm{in}}_{l}(\mathbf{r};k_n) + \sum_{ll^\prime} ~i_{l}~S_{ll^\prime}(k_n) ~\psi^{\mathrm{out}}_{l^\prime}(\mathbf{r};k_n) \right] \\ \nonumber
&=& ~\left[ \sum_{l}~i_{l} ~\psi^{\mathrm{in}}_{l}(\mathbf{r};k_n) + \sum_{l^\prime} \alpha_{l^\prime} ~\psi^{\mathrm{out}}_{l^\prime}(\mathbf{r};k_n) \right] \\ \nonumber
\nonumber
			&\approx& ~\sum_{l^\prime}~\alpha_{l^\prime}~\psi^{\mathrm{out}}_{l^\prime}(\mathbf{r};k_n)
\end{eqnarray}
where we use in the last step that the scattering matrix $S_{ll^\prime}(k_n)$ is singular and thus neglect the first term, as explained above. With the singular vectors $\vec i$ and $\vec a$ we have thus defined scattering states which consist of purely outgoing waves. Equivalently to the fact that resonances of quantum maps do not only have right, but also left eigenstates, a resonance in the 3-disk system also corresponds to a scattering state of purely incoming waves. As $\overline{\psi}^{\mathrm{out}}_{l}(\mathbf{r};k) = \psi^{\mathrm{in}}_{-l}(\mathbf{r};\overline {k})$, this state is simply obtained by taking the complex conjugate of $\psi_n (\mathbf{r})$ .

\begin{figure}
  \raisebox{0.5cm}[0pt][0pt]{
  \includegraphics[width=.48\columnwidth]{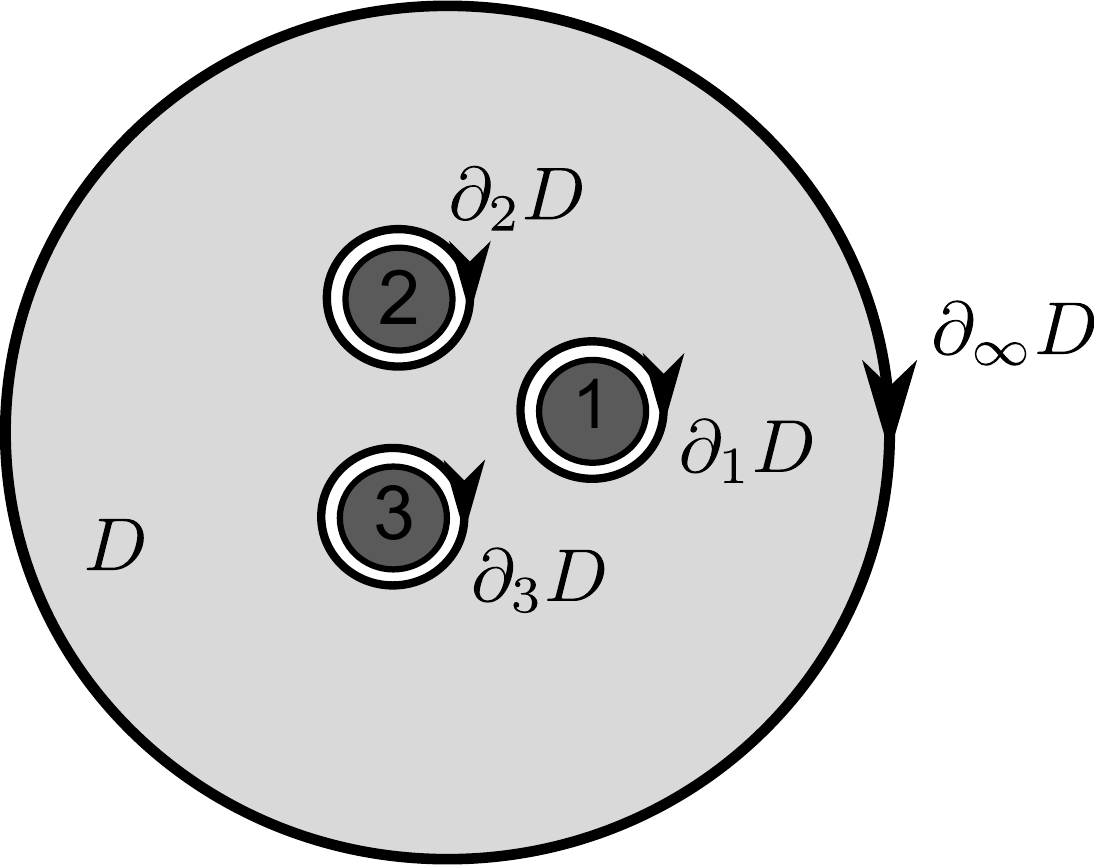}}
  \includegraphics[width=.42\columnwidth]{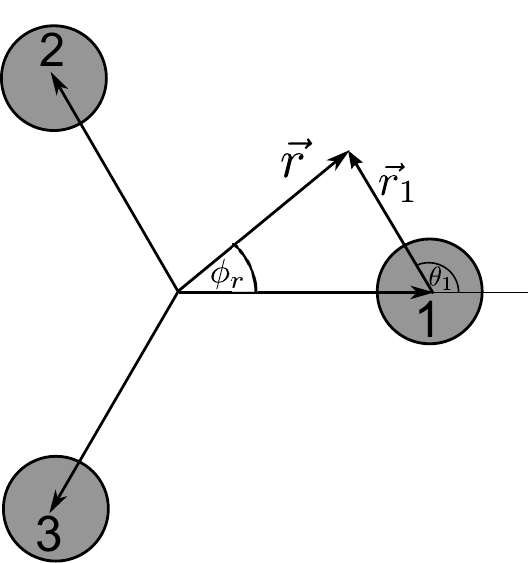}
\caption{\label{fig:def_Volume}
Left panel: Schematic drawing of the 3 disks (dark gray), the volume $D$ in light gray and its borders $\partial_i D$ very close around the disks and $\partial_\infty D$ limiting the volume at a large distance from the scattering center. Right panel: visualization of the different polar coordinate systems occurring in the calculations of the scattering states.}
\end{figure}

In order to implement this strategy, we require explicit knowledge of the scattering matrix. Based on Green's theorem Gaspard and Rice have shown that the wave function can be calculated via four surface integrals
\begin{equation}\label{eq:psi_integrals}
\psi_{l} (\mathbf{r};k) = I_\infty+\sum_{j = 1}^3 I_{lj}
\end{equation}
along the boundary of volume $D$, see \fref{fig:def_Volume}, and give explicit formulas for them in terms of Bessel and Hankel functions (see equations A11 and A6 in \cite{gas89d})
\begin{eqnarray} \label{eq:I_infty}
&I&_\infty (\mathbf{r}) = \exp(\rmi l\phi_r)J_l(kr) \\ \label{eq:I_j}
&I&_{lj} = -\frac{\pi a}{2\rmi} \sum_{m = -\infty}^\infty A_{ljm} ~H_m^{(1)}(k r_j)J_m(k a)\exp(\rmi m\theta_j)
\end{eqnarray}
for a position $(r_j,\theta_j)$ in the local polar coordinate system of disk $j$, see right panel in \fref{fig:def_Volume}. Here only the Fourier coefficients of the so called boundary wave function, which is the normal gradient of the wave function at the disk boundaries, occur. They are defined by
\begin{equation}\label{eq:WF_Fourier}
\mathbf{n}_j \cdot \nabla \psi_{l}(\mathbf{r}_j ;k) = \sum_{m = -\infty}^\infty A_{ljm} \exp(\rmi m\theta_j)
\end{equation}
where $j = 1,2,3$ are the labels of the disks and $\mathbf{r}_j$ is a point on the boundary of disk $j$ with polar angle $\theta_j$. Gaspard and Rice used \eref{eq:psi_integrals}, \eref{eq:I_infty} and \eref{eq:I_j} in order to derive the following expression for the scattering matrix
\begin{equation}\label{eq:SCMD}
 S_{ll^\prime} = 1-\rmi~\sum\limits_{j j^\prime m m^\prime}~C_{lj^\prime m^\prime} ~M^{-1}_{j^\prime m^\prime jm} D_{l^\prime j m}.
\end{equation}
The matrices $D,M$ and $C$ have explicit formulas in terms of Bessel and Hankel functions which can be found in \cite{gas89d}, and their connection to the expansion coefficients in \eref{eq:I_infty} is given by
\[
 A_{ljm} = \sum_{j^\prime m^\prime}~C_{lj^\prime m^\prime} ~M^{-1}_{j^\prime m^\prime jm}.
\]
From here all the ingredients of calculating the scattering state $\psi_n (\mathbf{r})$ are known: First find the singular vectors $\vec i$ and $\vec a$ of $S$ according to \eref{eq:singular_S}, then use \eref{eq:psi_scat} and find the final formula
\begin{eqnarray} \label{eq:Psi_Scat_Calc}
\psi_n (\mathbf{r}) &=& \sum_{l} i_{l} ~\left( I_\infty+\sum_{j = 1}^3 I_{lj}\right) \\ \nonumber
 &=&-\frac{\pi a}{2\rmi} \sum_{j,m} \left(\sum_{l} i_{l} A_{ljm}\right) H_m^{(1)}(k_nr_j)J_m(k_na)\exp(\rmi m\theta_j),
 \\ \nonumber
 &=&-\frac{\pi a}{2\rmi} \sum_{j,m} A_{jm;n} H_m^{(1)}(k_nr_j)J_m(k_na)\exp(\rmi m\theta_j),
\end{eqnarray}
where we defined
\begin{equation} \label{eq:Ascat_def}
A_{jm;n} :=\sum_{l} i_{l} A_{ljm}.
\end{equation}
The $I_\infty$-term is neglected because we will see in \eref{eq:def_v} and \eref{eq:av} that it is small compared to the terms containing $A_{jm;n}$. We now observe that using the decomposition \eref{eq:SCMD}, it is possible to access the $A_{jm;n}$ directly via the matrix $M$. This parallels the considerations of Gaspard and Rice, who obtained the resonances $k_n$ as roots of the $M$-matrix instead of finding the poles of $S$. Analogously, we can obtain the resonance states from the singular vectors $\vec w$ and $\vec v$ of $M$ instead of looking for the singular vector $\vec i$ of $S$. These singular vectors are obtained by a standard singular value decomposition such that
\begin{equation} \label{eq:def_v}
 \sum\limits_{j^\prime m^\prime}w_{j^\prime m^\prime} ~M^{-1}_{j^\prime m^\prime j m } = \infty~ v_{jm}.
\end{equation}
The comparison with \eref{eq:SCMD} shows that the singular vector of $S$ must be related to the singular vector of $M$ via
\begin{equation}\label{eq:def_i}
\sum_{l}i_{l}~C_{lj^\prime m^\prime} = w_{j^\prime m^\prime}.
\end{equation}
Using this relation and the equality $A= CM^{-1}$ we find based on \eref{eq:Ascat_def}
\begin{eqnarray} \nonumber
A_{jm;n} &=& \sum_{l} i_{l} A_{ljm} \\ \nonumber
	&=& \sum_{j^\prime m^\prime}~\sum_l~i_l~C_{lj^\prime m^\prime} ~M^{-1}_{j^\prime m^\prime jm} \\ \nonumber
 &=& \sum_{j^\prime m^\prime}~w_{j^\prime m^\prime } ~M^{-1}_{j^\prime m^\prime jm} \\
 &=& \infty v_{jm}.
 \label{eq:av}
\end{eqnarray}
Summarizing these results we have shown that we can calculate scattering states knowing only their Fourier coefficients on the disks boundary. These can be obtained directly by a singular value decomposition of the matrix $M$. This procedure is preferable in numerical calculations due to its speed and its stability compared to going the way of decomposing the complete scattering matrix. In the next section we will see that these Fourier coefficients also provide an efficient way to calculate phase space representations of the resonant scattering states.

\subsection{Calculation of phase space representations}
\label{sec:PhaseSpaceRep}

If one is interested in the relation between the resonances and the underlying classical system it is necessary to study the resonance states not only in configuration space, but also in phase space. For closed systems Wigner \cite{wig32} or the Husimi functions \cite{hus40} can be used for this purpose. For open systems, where each resonance corresponds to two different states (left and right resonance states, corresponding to incoming and outgoing boundary conditions) Ermann, Carlos and Saraceno proposed an additional phase space representation which takes both kinds of states into account \cite{erm09}. To visualize such representations for two-dimensional billiards it is very common to reduce the three-dimensional energy shell to the canonical two-dimensional Poincar\'{e} section of boundary reflections. In this section we first recall the definition of the conventional Poincar\'{e}-Husimi distributions and then show how these can be efficiently calculated in the $S$-matrix approach. Finally we apply the idea of Ermann, Carlo and Saraceno to the Poincar\'{e}-Husimi distributions and obtain a symmetric phase space distribution, which we will call \emph{ECS distribution} throughout this paper.

Let us start with the definition of the Poincar\'{e}-Husimi distribution. Following \cite{cre93,baec04},
this distribution is obtained from the boundary function
\[
u_n(s) := \hat \mathbf{n}(s) \cdot \nabla \psi_n(\mathbf{r}(s)),
\]
where $\mathbf{r}(s)$ is a point on the boundary of one of the disks, parametrized by the dimensionless arclength $s\in[-\pi,\pi]$ (measured again in units of $a$),
and $\hat \mathbf{n}(s)$ is the normal vector at this point. The Poincar\'{e}-Husimi distribution
\begin{eqnarray} \label{eq:DefPoincHusimi}
H_n(q,p) &\propto& |\langle c_{(q,p),\mathrm{Re}\,k_n}| u_n\rangle|^2\\ \nonumber
  &=&\left| \int_{\partial_i D} \overline{c}_{(q,p),\mathrm{Re}\,k_n}(s) u_n(s) \rmd s\right|^2
\end{eqnarray}
then follows by a projection onto a coherent state
\[
c_{(q,p),k}(s) \propto \sum_{l\in Z}\rme^{\rmi ka p(s-q+2\pi l)-\frac{ka}{2}(s-q+2\pi l)^2},
\]
where $(q,p)$ are the Birkhoff coordinates on the disk. The Poincar\'{e}-Husimi function can be normalized afterwards, but in contrast to closed systems the normalization factor cannot be given in closed form.

Using the Fourier decomposition \eref{eq:WF_Fourier} for the boundary function we obtain at the boundary of disk $j$
\begin{eqnarray} \fl \nonumber
H_n(q,p) &\propto&\Big| \int_0^{2\pi} \sum_{l\in \mathbb{Z}}\rme^{-\rmi p\,\mathrm{Re}\,k_n a(s-q+2\pi l)-\frac{\mathrm{Re}\,k_n a}{2}(s-q+2\pi l)^2} \sum_{m \in \mathbb{Z}} A_{jm;n} \rme^{\rmi m s} \rmd s\Big|^2 \\ \fl\nonumber
&=&\left| \sum_{m \in \mathbb{Z}} A_{jm;n} \int_\mathbb{R} \rme^{\rmi[m-p\,\mathrm{Re}\,k_n a]s} \rme^{-\frac{\mathrm{Re}\,k_n a}{2}(s-q)^2} \rmd s\right|^2,
\end{eqnarray}
which can be further evaluated to finally give
\begin{eqnarray} \fl\nonumber
H_n(q,p) &\propto& \left| \sum_{m \in \mathbb{Z}} A_{jm;n} \rme^{\rmi mq} \rme^{-(p\,\mathrm{Re}\,k_n a-m)^2/(2\,\mathrm{Re}\,k_n a)} \right|^2.
\end{eqnarray}
Again only the Fourier coefficients $A_{jm;n}=v_{jm}$ of the resonance state appear in the expression, which thus can be evaluated knowing the singular vector $\vec v$. In contrast to \eref{eq:Psi_Scat_Calc} for the wave function, the formula for the Poincar\'{e}-Husimi distribution contains no Bessel and Hankel functions but only exponential functions, and can thus be calculated much faster.

The simple but efficient idea of Ermann, Carlo and Saraceno is to take not only the left or the right resonant states for the definition of the Husimi distribution, but to take the product of the projection of both states on the coherent state \cite{erm09}. Applied to the definition of the Poincar\'{e}-Husimi representation \eref{eq:DefPoincHusimi} this leads to
\[
 H^{ECS}_n(q,p) \propto \langle c_{(q,p),k_n}| u^R_n\rangle\langle u^L_n|c_{(q,p),k_n}\rangle
\]
where $u^L_n$ and $u^R_n$ are the boundary functions of the left and right resonance states, respectively.
Since these states are related by complex conjugation, the procedure described above directly carries over to this representation.

\section{Numerical results}
\label{sec:results}

Via the resonance wave functions and the associated Poincar\'{e}-Husimi and ECS distributions we are now able to study the properties of the resonance states in the 3-disk system both in configuration space and in phase space. We first verify the observations from quantum maps that long-living and short-living states localize on and off the classical trapped sets, respectively. We then turn to the correlations of these properties along the resonance chains. This allows us to identifying a pairwise interaction of chains driven by the interplay of the fundamental periodic orbits. Furthermore, we gain insight into the gradual merging with additional chains as the phase-space resolution of the trapped sets increases in the semiclassical limit of large $\mathrm{Re}\,k_n$. This illuminates how the key mechanism behind the fractal Weyl law in quantum maps extends to autonomous systems.

\subsection{Localization on classical phase space structures}
\label{sec:loc}

\begin{figure}\centering
  \includegraphics[width=0.98\textwidth]{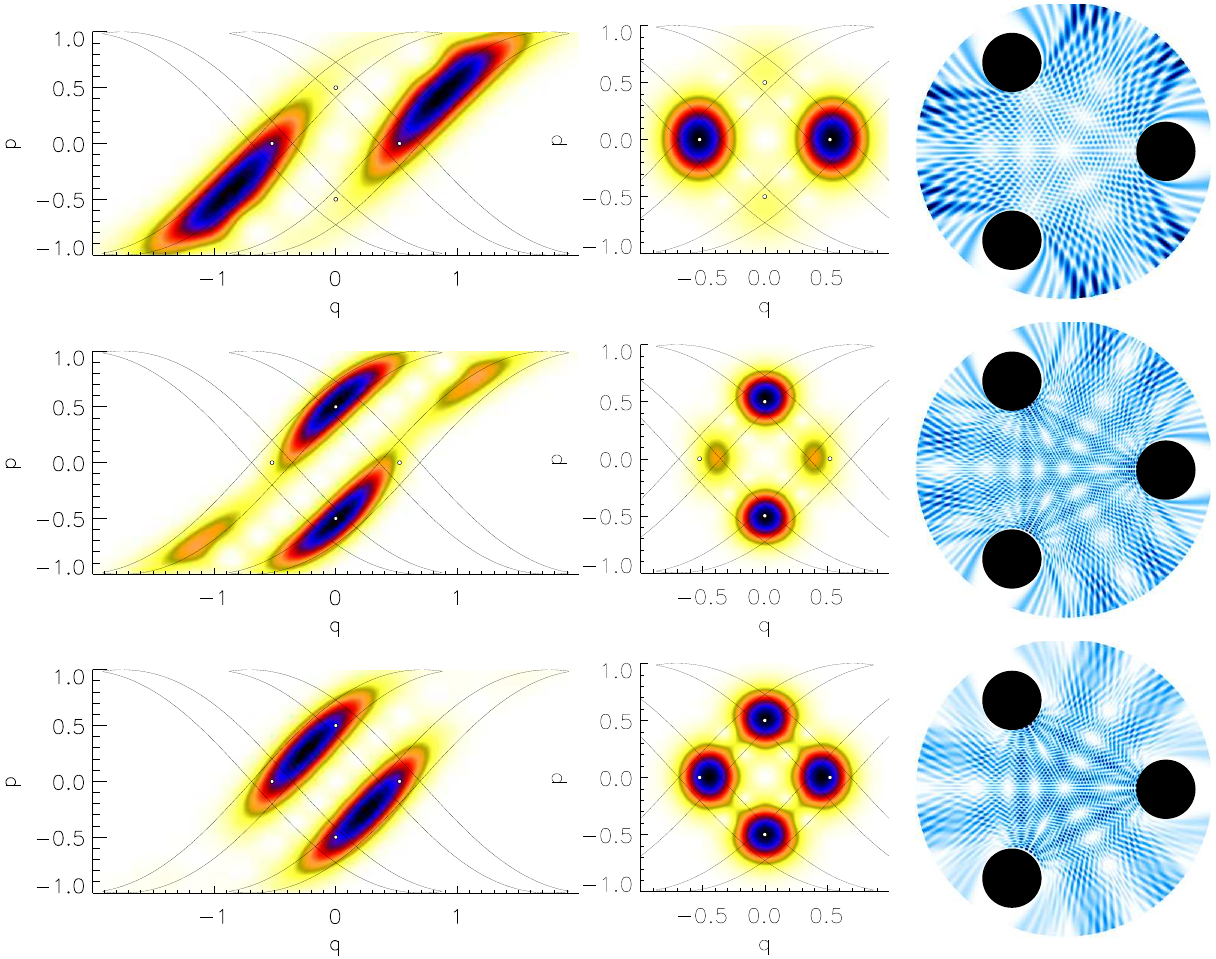}
\caption{\label{fig:loc_Orb0}
Visualizations of three long-living resonance states (indicated in \fref{fig:Res_chains} by the black crosses on the red chain). Left: Husimi distribution in a color code where white denotes vanishing density and black denotes maximal density. Also indicated are the boundaries of the trapped set of order 1 (black lines) and the positions of the fundamental orbits $0$ and $1$ (dots). Centre: Analogous plot of the ECS distribution. Right: Intensity of wave function in configuration space in the color code ranging from white to blue, with the 3 disks plotted in black. The top panels show the resonance state at $k_na = 23.66-0.26\rmi$, which displays localization on orbit $0$. The middle panels show the resonance at $k_na = 43.47-0.20\rmi$, which is localized on orbit $1$. The bottom panels refer to the resonance at $k_na = 35.99-0.15\rmi$; in this case, the state is localized on both orbits.}
\end{figure}

\Fref{fig:loc_Orb0} shows three examples of long-living resonance states. The corresponding resonances are marked by black crosses on the red chain in \fref{fig:Res_chains}. The top panel shows an example of localization around the orbit $0$. This is already seen in the Poincar\'{e}-Husimi distribution, but comes out prominently in the ECS distribution, where the intensity accumulates on the region $\mathcal T^{(1)}_0$ of the trapped set. We also note that the wave function shows a dominant structure around the bouncing ball orbit $0$, and a nodal line around orbit $1$.
In contrast, the wave function in the middle panels in \fref{fig:loc_Orb0} shows a clear enhancement on the triangular orbit $1$, which is confirmed by the phase space localization in the Poincar\'{e}-Husimi and ECS distributions. Note that in \fref{fig:Res_chains}, both of these states are near a crossing of the red chain with the blue chain. The bottom panels in \fref{fig:loc_Orb0} illustrate the situation away from these crossings. In the ECS distribution, the resonance states now localize on both fundamental orbits, while in the conventional Poincar\'{e}-Husimi distribution also occupies the region between these orbits on the backward trapped set. The wave function shows an enhanced intensity on both fundamental orbits.

\begin{figure}
  \centering
  \includegraphics[width=0.98\textwidth]{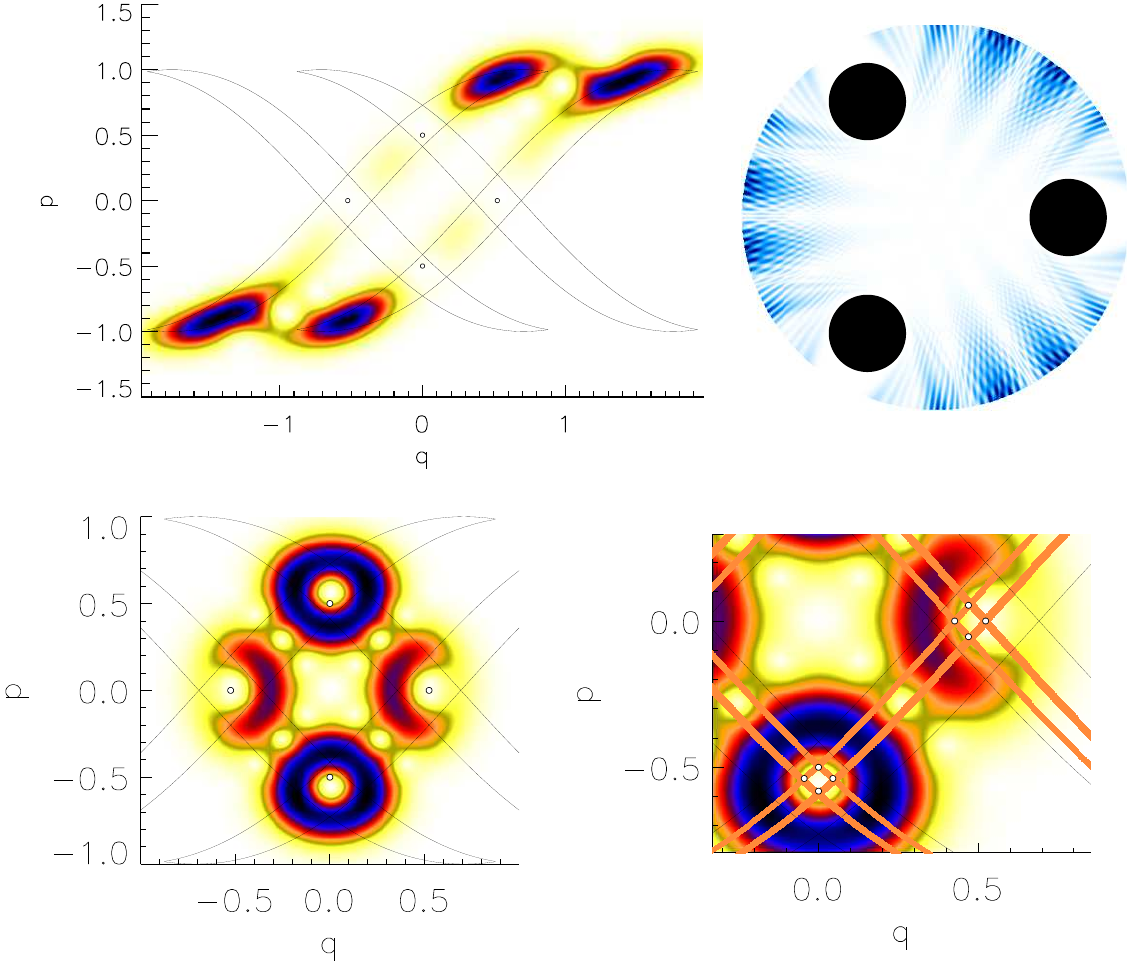}
\caption{\label{fig:deepRes}
Same as in \fref{fig:loc_Orb0}, but for the short-living resonance at $k_na = 54.87-0.71\rmi$, and with an additional panel (bottom right) showing a close-up of the ECS distribution (now placed bottom left). The phase space distributions now avoid the trapped set. In the close-up, the orange areas are the forward and backward trapped sets of order 2, while the black circles indicate the position of some higher-order periodic orbits.}
\end{figure}

In contrast, the short living scattering states, lying deeper in the complex $k$-plane, avoid the regions around the periodic orbits. A typical example is given in \fref{fig:deepRes}. Here the wave function lives outside the scattering region and locates on the cusps of the backward trapped set in the Poincar\'{e}-Husimi representation.
The ECS distribution forms rings or half rings around the trapped set.

As expected, the resonance states for the 3-disk system show the same localization behavior (on and off the trapped set for long living and short living resonances) as it has already been reported for quantum maps. However, in quantum maps these resonances do not organise into chains. We thus now proceed to the analysis of this additional aspect.

\subsection{Interaction and correlations of resonance chains}
\label{sec:ResChains}

Looking at the resonances in \fref{fig:Res_chains}, the chain structure is the most striking feature. As discussed in the introduction, such resonance chains are a common feature of open quantum mechanical systems with ballistic classical mechanics, or analogous wave-optical systems. While these chains are commonly associated with classical orbits, we now show that one can develop a much more detailed understanding by considering the interactions of the chains, and complement these with the correlations of the resonance states along the chains.

\begin{figure}
  \centering
  \includegraphics[width=0.85\textwidth]{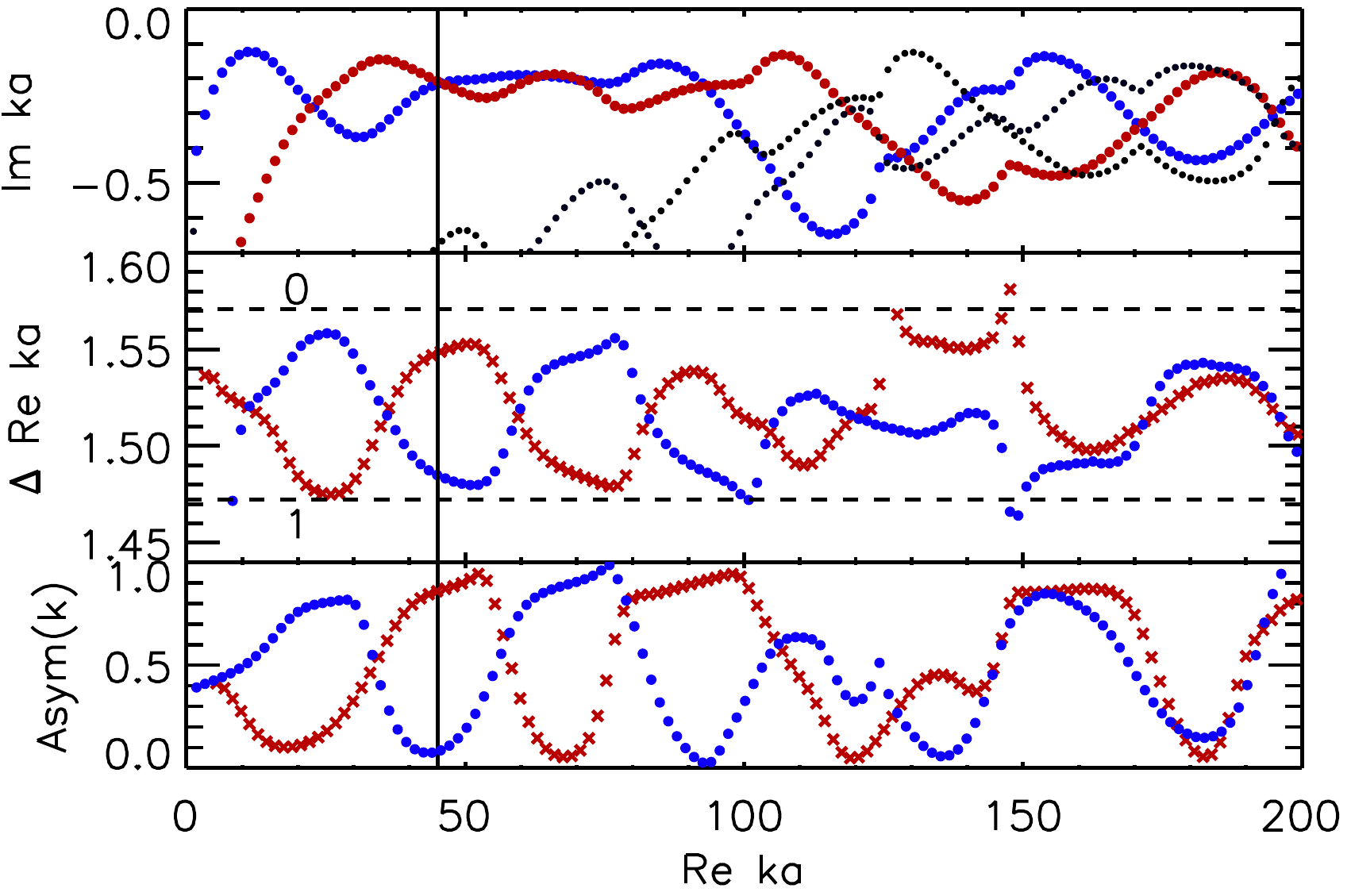}
\caption{\label{fig:Asym}
Resonance spacings $\Delta\,\mathrm{Re}\,ka$ (middle panel) and the phase-space asymmetry defined in \eref{eq:asym} (lower panel) for the resonances of the two chains of long-living resonances (highlighted in blue and red in the upper panel). The vertical line marks the position of a crossing near $\mathrm{Re}\,k_na = 46$. The two horizontal dashed lines in the middle panel indicate the expected resonance spacing for a WKB quantization of the two fundamental orbits $0$ and $1$.}
\end{figure}

\begin{figure}
  \centering
  \includegraphics[width=0.45\textwidth]{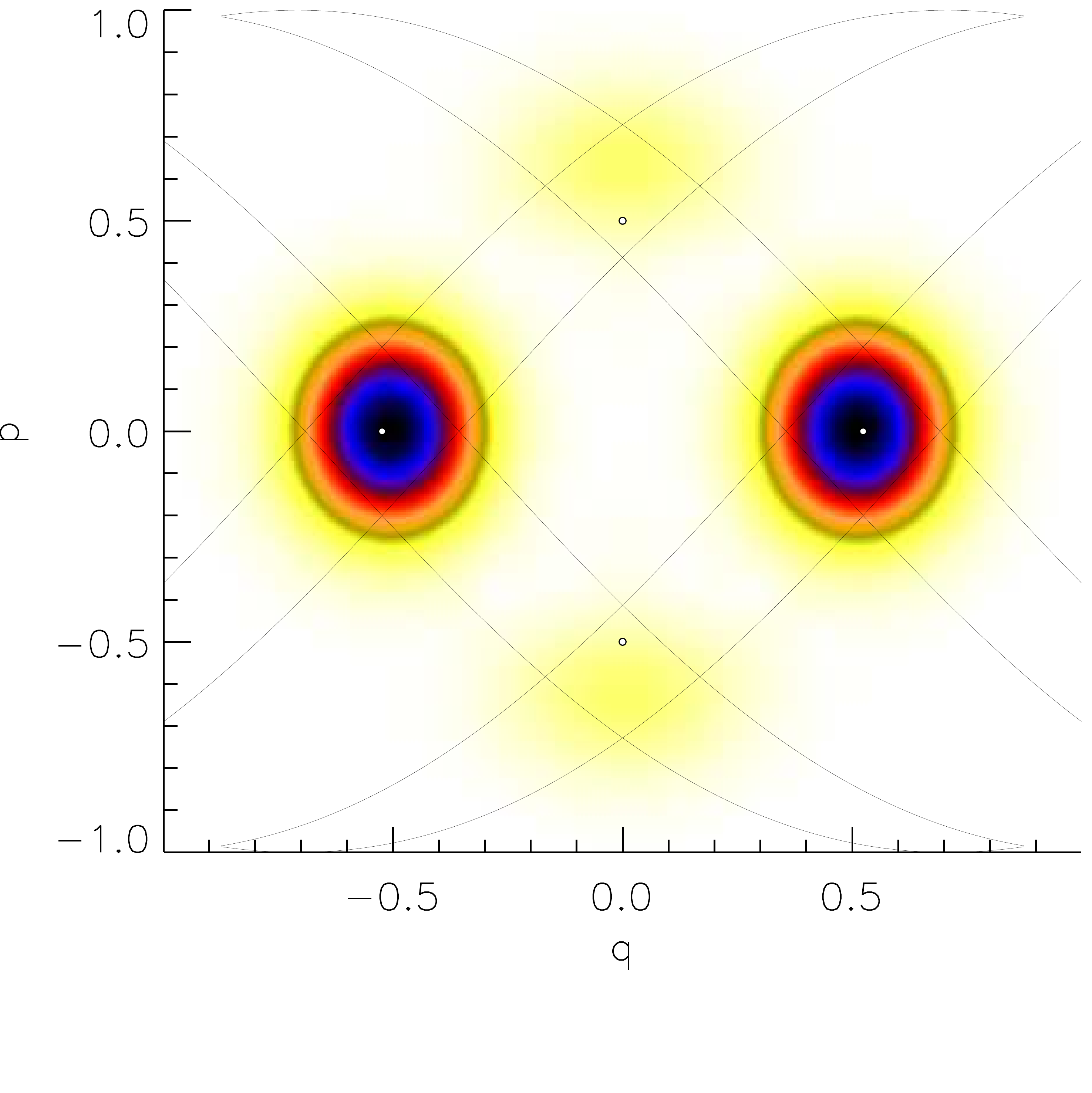}
  \includegraphics[width=0.45\textwidth]{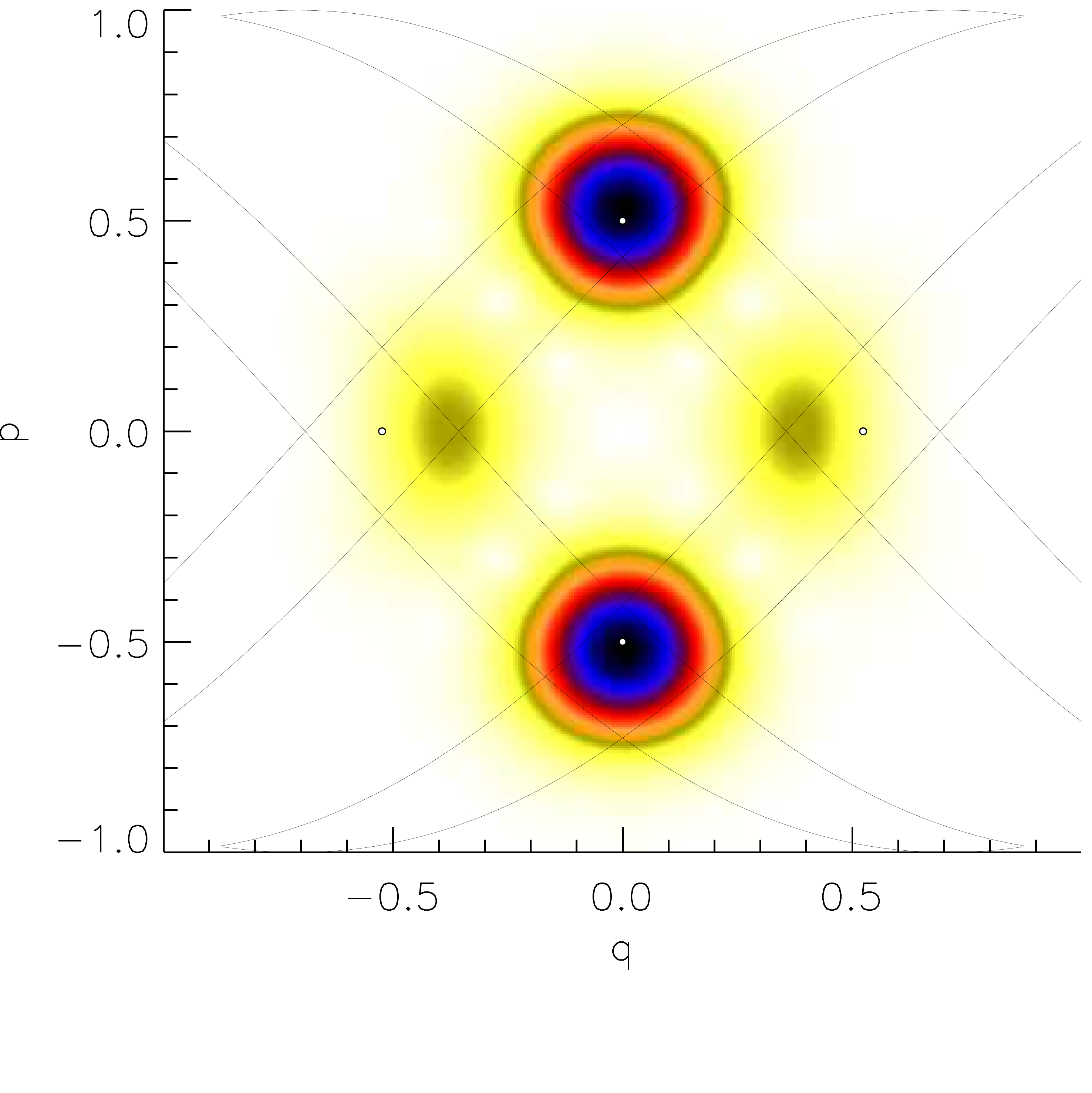}
\caption{\label{fig:crossRes}
ECS distribution of resonances at $k_na = 45.60-0.21\rmi$ and $k_na = 46.44-0.22\rmi$, sitting closest to the crossing of resonance chains indicated by the vertical line in \fref{fig:Asym}.}
\end{figure}

The interactions occur in the form of systematic crossings of pairs of chains that are otherwise well separated.
We focus on the two chains of long-living resonances (shown in more detail in the top panel of \fref{fig:Asym}), which display a systematic pattern of crossings in the region of $\mathrm{Re}\,k_na< 100$, while a merging with the chains of originally short-living resonances is observed at larger $\mathrm{Re}\,k_na$. Since we consider the spectrum of the $A_2$ symmetry-reduced 3-disk system, there exist no further symmetries that could be responsible for the crossings of chains.

We already have seen that the states along these chains of long-living states localize in the regions around the fundamental orbits (see \fref{fig:loc_Orb0}). One reason why the visual impression of these chains is that strong is the fact that the spacing $\Delta \mathrm{Re}\,ka$ between consecutive resonances in each chain is almost constant, and approximately equal to $1.5$ (see the middle panel of \fref{fig:Asym}). This value corresponds to a length of $\pi a/1.5 = 2.094\,a$, which has been observed \cite{cvi97,wir99b} to coincide well with half the length $L_0/2=2a$ and $L_1/2=2.13a$ of the fundamental periodic orbits in the symmetry-reduced system. However, \fref{fig:loc_Orb0} already suggests that the chains do not simply arise from a WKB-quantization of one of these orbits. The three resonance states shown there belong to the same (red) chain, but localize either on orbit 0, on orbit 1, or on both orbits. Thus it is not possible to associate one chain to a single orbit only.

A more detailed inspection of the resonance spacings in \fref{fig:Asym} supports this observation. The spacings generally do not coincide exactly with the spacing $k_0a=2\pi a/L_0=1.57$ and $k_1a=2\pi a/L_1=1.47$ expected from the fundamental orbits, but oscillate between these two values, which are indicated by the dashed horizontal lines. Only close to the crossings of the red and blue chains do the spacings approach $k_0$ and $k_1$. An example of this is the situation at the position of the vertical black line. The ECS distributions of the states on each chain that are closest to the crossings are shown in \fref{fig:crossRes}. As already seen in the top and middle panels of \fref{fig:loc_Orb0}, such states are strongly localized on the regions around a fundamental orbit, but now we can confirm that the orbit is selected corresponding to the observed resonance spacing.

At positions where the resonance chains separate from each other, the resonance spacings take values in between $k_0$ and $k_1$. This agrees with the localization pattern observed in the bottom panel of \fref{fig:loc_Orb0}. As a matter of fact, we see that the spacings themselves form chains and display systematic crossings. This explains why states on the same chain can be localized on either of the two fundamental orbits (see again top and middle panels of \fref{fig:loc_Orb0}), and indeed suggests the existence of systematic oscillations in the localization pattern.

\begin{figure}
  \centering
  \includegraphics[width=0.8\textwidth]{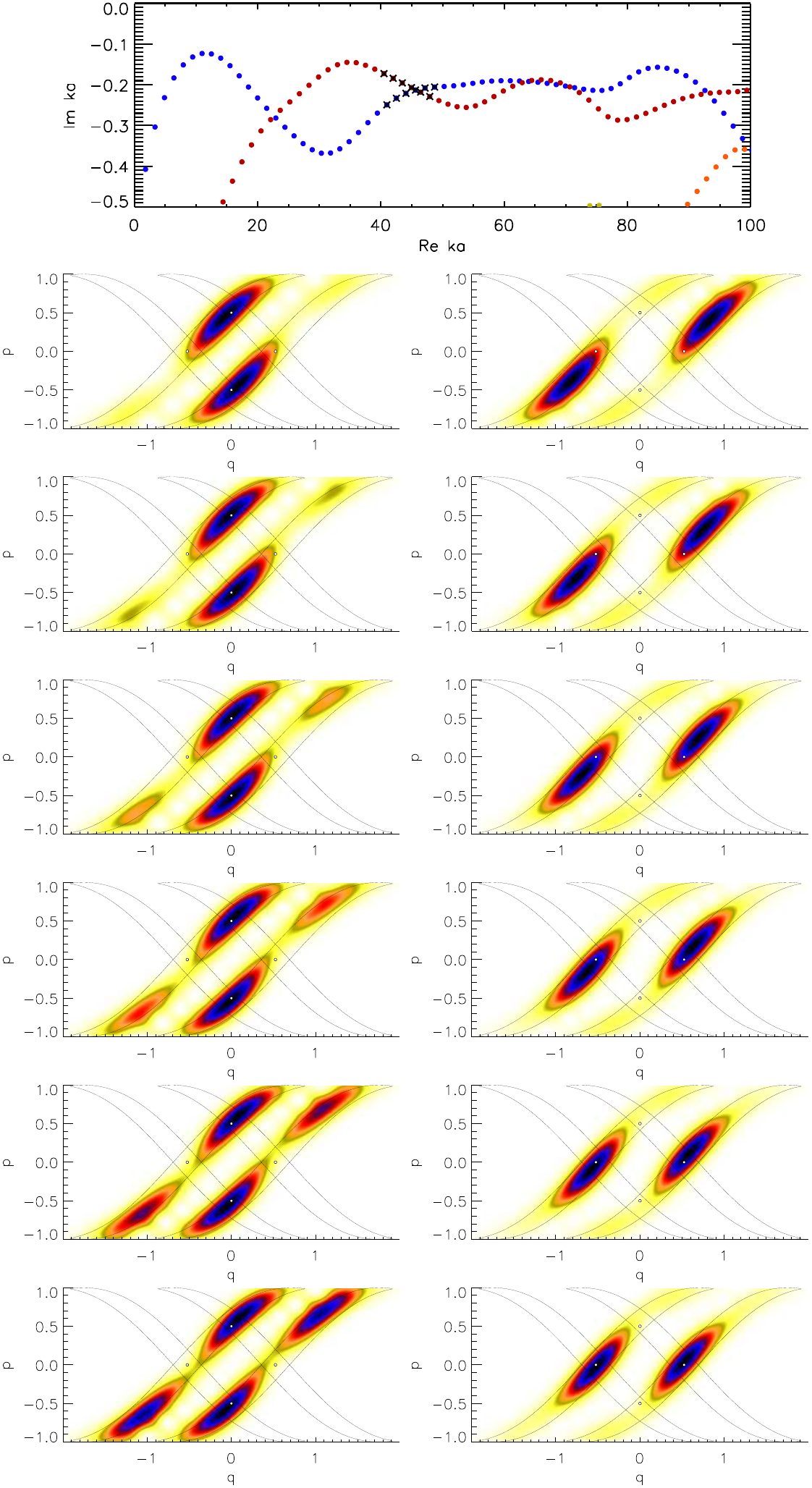}
\caption{\label{fig:Res_evol}
Poincar\'{e}-Husimi distributions of consecutive resonances belonging to the red chain (left column) and the blue chain (right column). These resonances are marked by the black crosses in the upper panel.}
\end{figure}

In order to quantify the localization on the fundamental orbits we introduce the quantity
\begin{equation}\label{eq:asym}
\mathrm{Asym}(k_n) = \frac{\int_{\mathcal T^{(1)}_0} |H_n^{\mathrm{ECS}}(q,p)|dqdp}{\int_{\mathcal T^{(1)}_0 \cup \mathcal T^{(1)}_1} |H_n^{\mathrm{ECS}}(q,p)|\rmd q\rmd p}.
\end{equation}
This measures the asymmetry of the projection of the ECS distribution onto the subsets $\mathcal T^{(1)}_1$ and $\mathcal T^{(1)}_0$ of the first order trapped set, associated to the orbits 0 and 1. The dependence of this quantity on $\mathrm{Re}\,k_na$ is shown in the bottom panel of \fref{fig:Asym}. As expected, near the crossing of resonance chains the asymmetry approaches the extremal values of 0 and 1, indicating complete localization on $\mathcal T^{(1)}_0$ or $\mathcal T^{(1)}_1$. In between these values the asymmetry indicates localization in both of these regions. Overall, the asymmetry displays oscillations that coincide very well with the oscillations of the resonance spacings. Similar oscillations are observed in the Poincar\'{e}-Husimi distribution. This is verified in \fref{fig:Res_evol}, which shows the Poincar\'{e}-Husimi distributions of consecutive resonances on both chains while moving along the chains.

The observed localization patterns suggest the existence of systematic correlations of the states of interacting chains. Resonances at similar $\mathrm{Re}\,ka$ belonging to different chains tend to have phase space representations which avoid each other; in \fref{fig:Res_evol}, this is manifested in a clockwise shift of the support on the backward trapped set as one moves along the resonance chain. On the other hand, a finite amount of correlations is expected as one compares resonances on interacting chains at different $\mathrm{Re}\,ka$.

\begin{figure}\centering
  \includegraphics[width=0.49\textwidth]{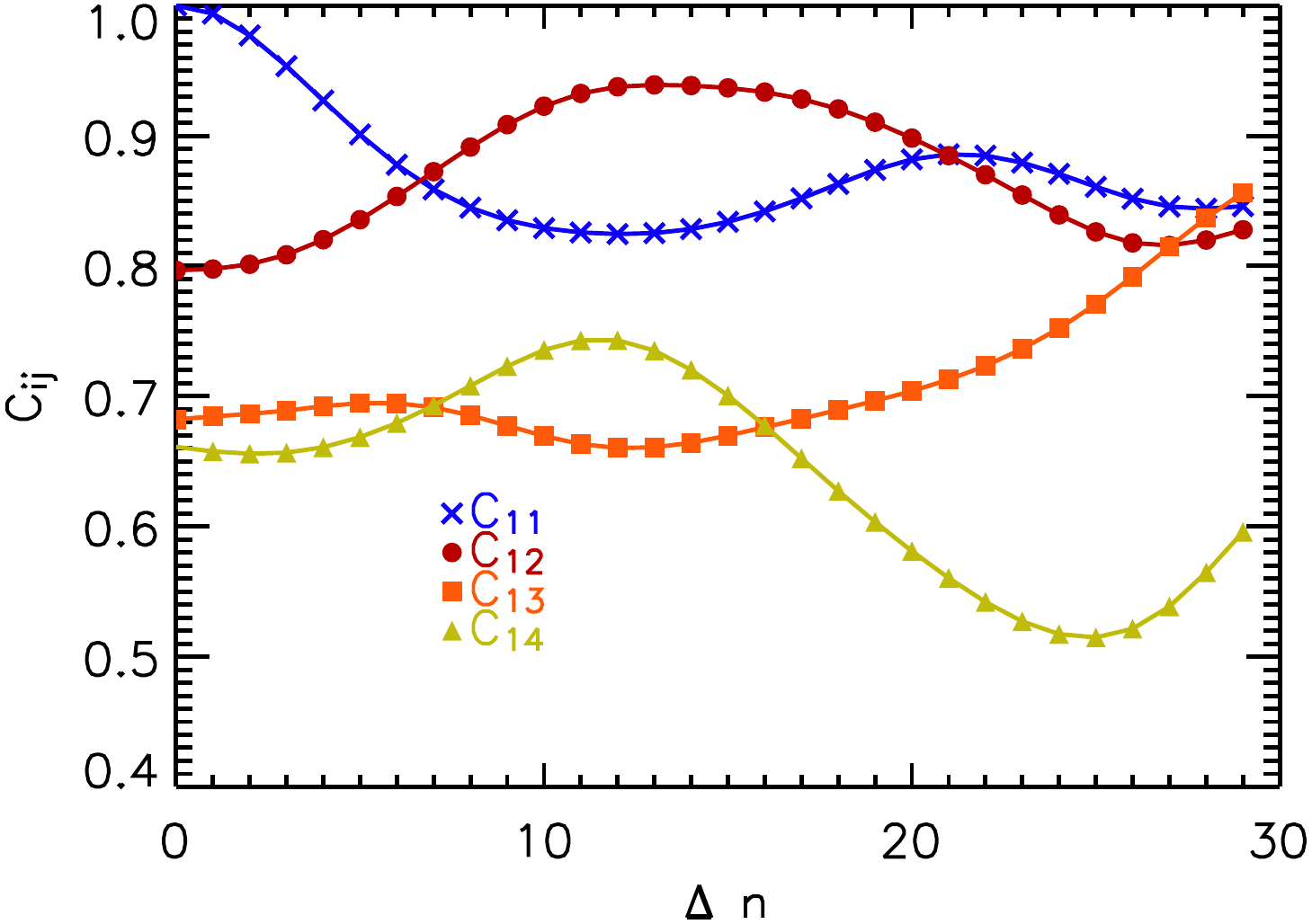}
  \includegraphics[width=0.49\textwidth]{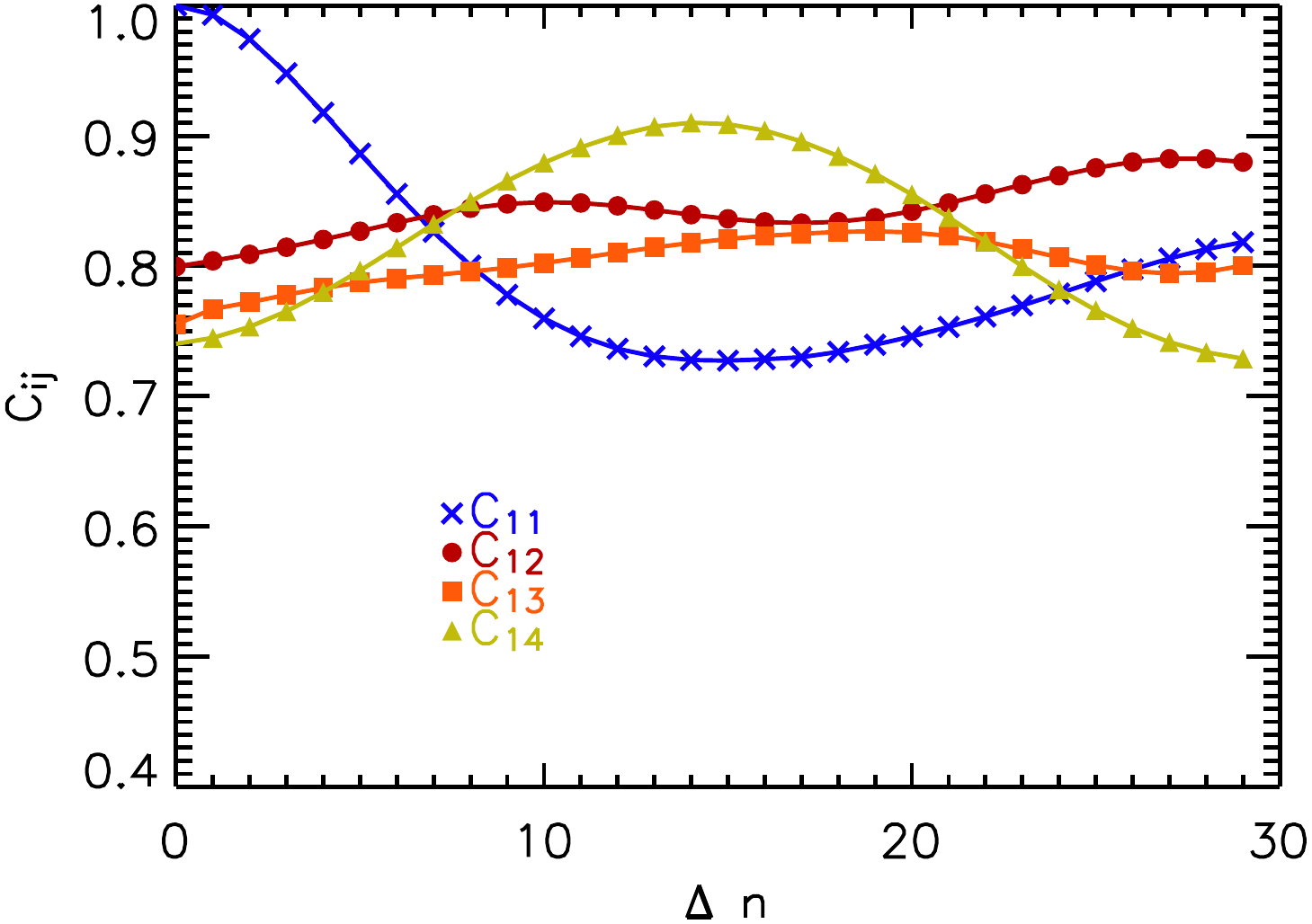}
\caption{\label{fig:Correl}
Correlation functions $C_{11}$, $C_{12}$, $C_{13}$ and $C_{14}$, defined in \eref{eq:Correlfunc}, calculated in the range $50 \le \mathrm{Re}\,ka \le 100$ (left) and $100 \le \mathrm{Re}\,ka \le 200$ (right).}
\end{figure}

To quantify such correlations we study the averaged pairwise overlap
\begin{equation}\label{eq:Correlfunc}\fl
C_{ij}(\Delta n) = \frac{1}{\mathrm{max}+1-\mathrm{min}}\sum_{n=\mathrm{min}}^{\mathrm{max}} \int \sqrt{ H_n^{(i)}(q,p)} \sqrt{H_{n+\Delta n}^{(j)}(q,p)} \rmd q\rmd p.
\end{equation}
of Poincar\'{e}-Husimi representations displaced by a fixed number $\Delta n$ of resonances along the chains. Here $i$ and $j$ label the different chains, so that the case $i=j$ corresponds to an autocorrelation function while $i\neq j$ signifies cross-correlations. We weighted the overlaps by taking the square root as the distributions already correspond to a probability density.

\begin{figure}
  \centering
  \includegraphics[width=0.9\textwidth]{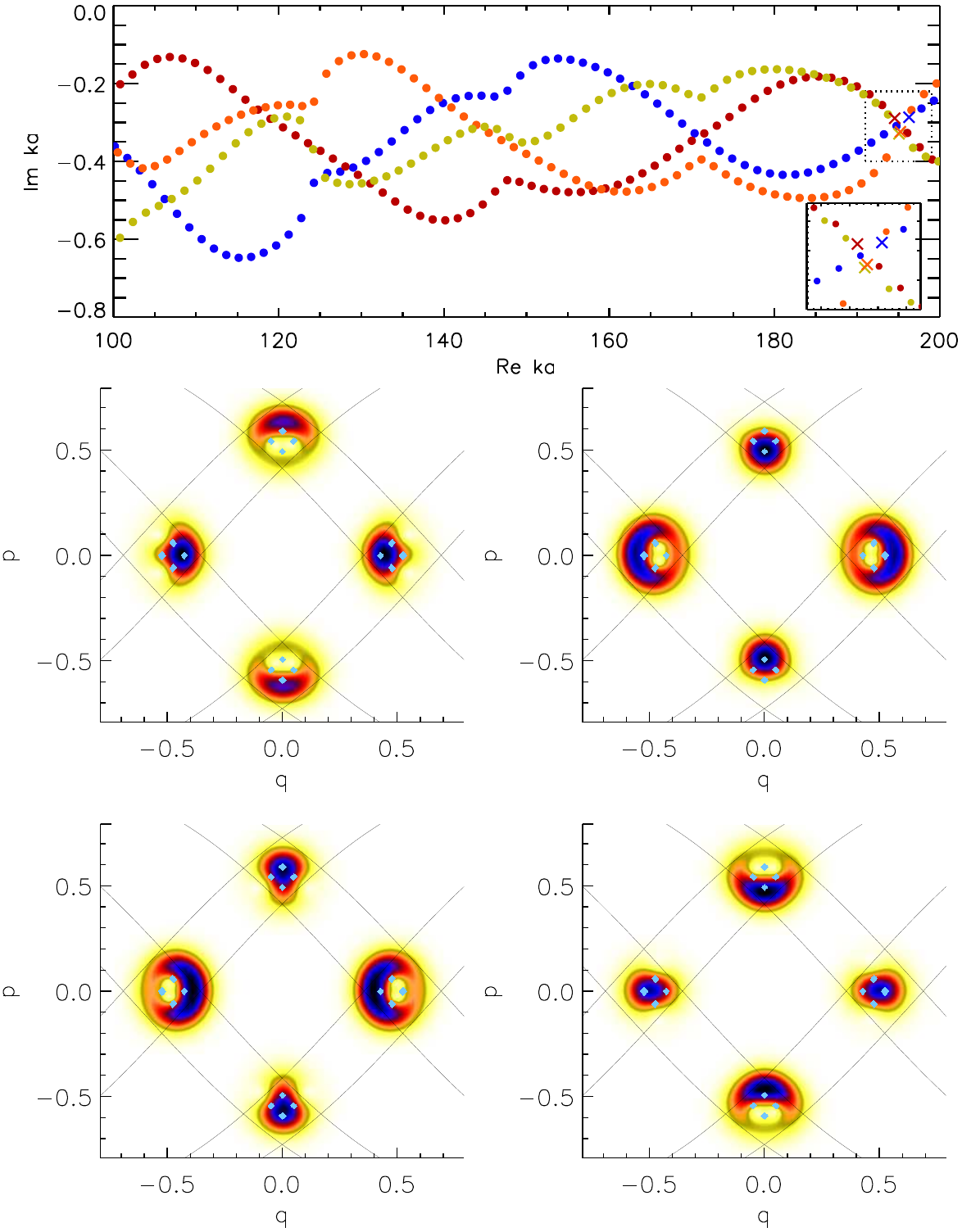}
\caption{\label{fig:HusimiMergedChains}
ECS distributions of the four nearby resonances at $\mathrm{Re}\,k_na \approx 195$ from four different chains (indicated by crosses in the upper panel, see in particular the zoom-in in the lower right corner). The black lines show again the borders of the forward and backward trapped set of order one. The light blue rectangles indicate the trapped set of order two. One can see that the four states fill the trapped set of order one complementarily and already bend around the structures of the order two trapped set.}
\end{figure}

\Fref{fig:Correl} shows in the left panel the correlation functions for resonances in the range $50 \le \mathrm{Re}\,ka \le 100$, where chain 1 (blue in \fref{fig:Res_chains}) interacts with chain 2 (red), but is well separated from chains 3 and 4 (orange and pale green, respectively). The autocorrelation function $C_{11}(\Delta n)$ starts at a maximum and then displays oscillations that correspond well to the oscillations in \fref{fig:Asym}, with the first minimum arising at the distance at which the localization pattern of the resonance states shifts between the two fundamental orbits. As expected, the cross-correlation function $C_{12}(\Delta n)$ with the interacting chain shows a complementary oscillatory behavior, while correlations $C_{13}$ and $C_{14}$ are small.

So far we focused our discussion on the pairwise interaction of the two chains of longest-living resonances,
which are well separated from the other chains for $\mathrm{Re}\,ka < 100$. However the fact that the level spacing along the chains is approximately constant together with the fractal Weyl law predicting an exponent of the counting function strictly greater than one implies that successively more and more chains have to merge from the short living to the long living regime. Such a first merging is observed around $\mathrm{Re}\,ka \approx 100$ from where on all four chains become entangled. Indeed, the interactions become very rich, and it is difficult to uniquely follow the chains through some of the crossings. As seen in \fref{fig:Asym}, both the resonance spacings as well as the asymmetry of the phase-space localization then display a more erratic behaviour. The effect on the correlation functions is illustrated in the right panel of \fref{fig:Correl}, which is obtained from the resonances in the range $100 \le \mathrm{Re}\,ka \le 200$. There is an indication of complementary oscillations in the auto-correlation function $C_{11}(\Delta n)$ and the cross-correlation function $C_{14}$, while in general the cross-correlations with all the chains are now of similar magnitude.

As the merging of different chains is responsible for the fractal Weyl law it should come along with a better resolution of the fractal trapped set and indeed \fref{fig:HusimiMergedChains} shows that in the regime where the four chains already merged the resonance states on all four chains populate the classical trapped set, and thus can be classified as long-living. Furthermore one observes that these distributions already bend around different regions of the second order trapped set, which is an indication that the phase space resolution at this frequency is already high enough to resolve the second order trapped set.

This increasing population of states on the classical trapped set has been previously observed in quantum maps, where this effect has been identified as the key mechanism behind the fractal Weyl law and our observations are a strong indication that this picture applies also to autonomous systems. However, since quantum maps generally do not display resonance chains, the detailed pathway via the interactions and merging of chains is a unique additional feature of autonomous systems.

\section{Quantum graph model of resonance chains}
\label{sec:quantumGraph}

In this section we present a quantum graph model which provides a phenomenological explanation of our numerical findings on interacting resonance chains in the 3-disk system, focussing on the regime of pairwise crossings for $\mathrm{Re}\,ka<100$. The model can be motivated by inspecting the semiclassical cycle expansion of the zeta function and therefore reproduces the salient features of the resonance spectrum. However, being based on a quantum graph it also allows to extract wave function information, which illuminates the origin of the semiclassical localization on the short periodic orbits and the $k$-dependent correlations between the resonance states.

The semiclassical zeta function provides a quantization condition based on the periodic orbits in the system \cite{cvi89,gas89a,kot99a}. In the cycle expansion, the contributions of these orbits are truncated in a manner which accounts for relations between long orbits and combinations of short orbits in the system.

The semiclassical zeta function of the $A_2$-symmetry reduced 3-disk system up to second order in the cycle expansion reads \cite{cvi89}
\begin{equation}\label{eq:quantCE}\fl
\zeta^{-1}(k)= 1+\frac{1}{\sqrt{\Lambda_0}}\rme^{\rmi kL_0} - \frac{1}{\sqrt{\Lambda_1}} \rme^{\rmi kL_1} + \bigg(\frac{\rme^{\rmi k(L_{01}-L_0-L_1)}}{\sqrt{\Lambda_{01}}}-\frac{1}{\sqrt{\Lambda_0 \Lambda_1}} \bigg) \rme^{\rmi k(L_0+L_1)} \, .
\end{equation}

\begin{figure}[t]
  \centering
  \raisebox{.6cm}{\raisebox{1cm}{
  \includegraphics[width=0.34\textwidth]{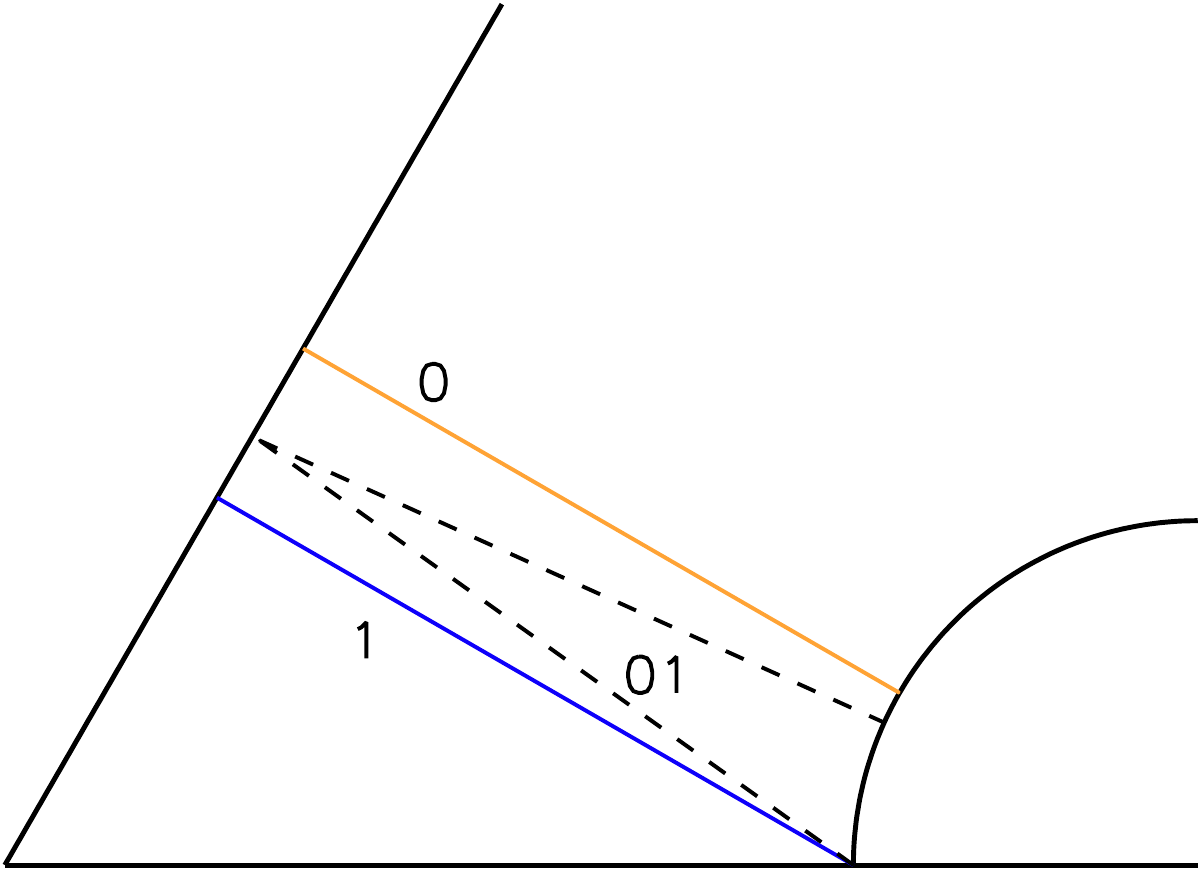}}}
  \includegraphics[width=0.64\columnwidth]{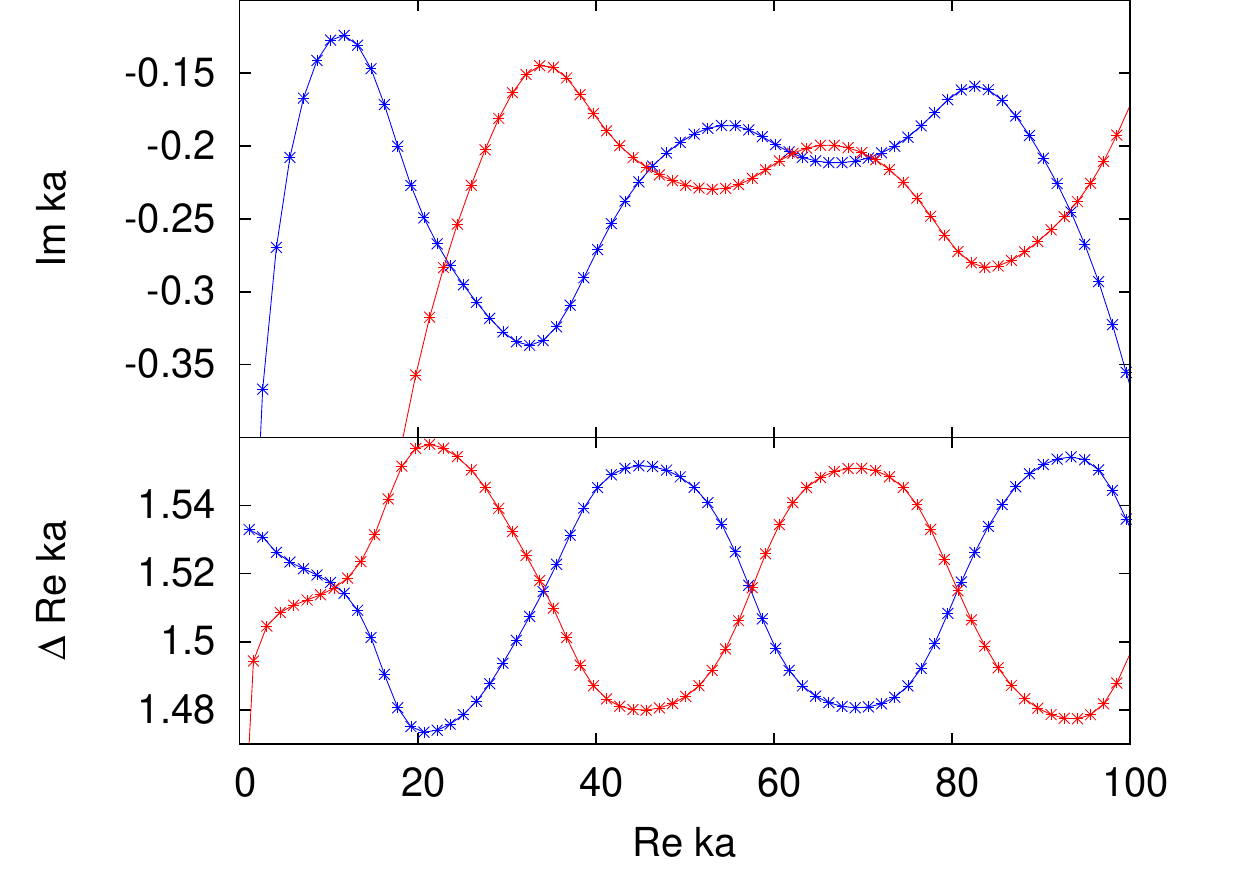}
\caption{\label{fig:figDisks}
The panel on the left shows a schematic representation of the fundamental domain of the 3-disk system ($R/a=6$) and the three shortest periodic orbits (solid lines: fundamental orbits $0$ (orange) and $1$ (blue); dashed line: their shadowing orbit $01$). Right: the semiclassical resonances obtained from the cycle expansion \eref{eq:quantCE} (top panel), as well as the corresponding resonance spacings $\Delta \mathrm{Re}\,k_n$ (bottom panel).}
\end{figure}

Here appear the three shortest primitive periodic orbits, the two fundamental ones denoted by 0 and 1 of first order and the only second order orbit denoted by 01. These orbits are illustrated in the left panel of \fref{fig:figDisks}, which displays the fundamental domain of a desymmetrised version of the system exploiting the $A_2$ symmetry. All properties of these orbits --~their lengths $L_0=4\,a$, $L_1=4.27\,a$, $L_{01}=8.32\,a$, and their stabilities $\Lambda_0=9.90$, $\Lambda_1=11.77$, and $\Lambda_{01}=124.09$~-- refer to this desymmetrised system. Note that $L_0+L_1\approx L_{01}$ and $\Lambda_0\Lambda_1\approx\Lambda_{01}$. The principle that the properties of longer orbits are well approximated by combinations of shorter ones is the central idea of the cycle expansion, leading to an extremely fast convergence \cite{cvi89}.

The zeros of $\zeta^{-1}(k)$ give the semiclassical resonances of the system, which are known \cite{cvi89,wir99b} to coincide well with the exact quantum mechanical ones except for the first few ones. In the right panel of \fref{fig:figDisks} we present the semiclassical resonances for the cycle expansion of second order and find two chains of resonances (which is connected to the fact that $L_0\approx L_1\approx L_{01}/2$; see the discussion at the end of this section). For $\mathrm{Re}\,ka<100$, where the two chains of longest-living resonances are isolated from the other resonance chains, we see, as already reported before \cite{cvi97,wir99b}, an excellent agreement (compare \fref{fig:Res_chains}). For larger $\mathrm{Re}\,ka$, where four chains interact with each others, higher orders in the cycle expansion have to be used (cf. discussion in \cite{cvi97,wir99b}).

\begin{figure}[b]
  \centering
  \includegraphics[width=0.6\textwidth]{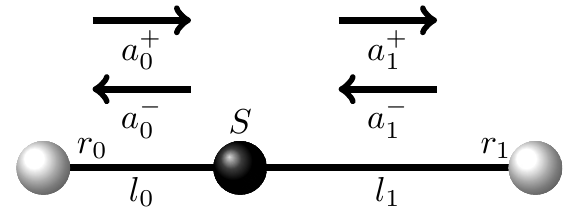}
\caption{\label{fig:1Dmodel}
Open quantum graph model, consisting of two coupled edges of length $l_0=L_0/2$ and $l_1=L_1/2$, leaky end nodes described by reflection coefficients $r_0$ and $r_1$, and a connecting node described by a coupling matrix $S$ \eref{eq:smat}. The coefficients $a_0^-$, $a_0^+$, $a_1^-$, and $a_1^+$ are the amplitudes of plane waves running up and down the two edges.}
\end{figure}

In quantum graphs, orbits of shorter length naturally combine into composite orbits of longer length \cite{kot99a}.
Thus, we consider a graph composed of two edges of length $l_0$ and $l_1$, on which the orbits 0 and 1 are localized (see \fref{fig:1Dmodel}), connected by a node which approximately combines these orbits into the longer orbit 01.
The nodes at the left and right ends of the graph describe leakage out of these orbits. Quantum-mechanically, these end nodes are described by reflection coefficients $r_0$ and $r_1$, while the node connecting both edges can be characterised by a $2\times 2$ coupling matrix $S$, which can be seen as a subblock of a scattering matrix which includes further leakage out of these periodic orbits. We parameterize this (nonunitary) matrix as
\begin{equation}\label{eq:smat}
S=\left(\begin{array}{cc}r & t \\
 t & r\end{array}\right)\, ,
\end{equation}
where $r$ and $t$ are complex numbers.

Each edge supports counter-propagating plane waves with wavefunction $\psi_p=a_p^+\exp(\rmi kx)+a_p^-\exp(-\rmi kx)$, with $p=0$ on edge 0 ($-l_0<x<0$) and $p=1$ on edge 1 ($0<x<l_1$). The matching conditions at the three nodes of the quantum graph read
\begin{eqnarray}\label{eq:coefWave}
  & & a_0^+\rme^{-\rmi kl_0} = r_0 a_0^- \rme^{\rmi kl_0}\, , \nonumber \\
  & & a_0^- = r a_0^+ + t a_1^- \, , \nonumber \\
  & & a_1^+ = t a_0^+ + r a_1^- \, , \nonumber \\
  & & a_1^- \rme^{-\rmi kl_1} = r_1 a_1^+ \rme^{\rmi kl_1} \, .
\end{eqnarray}
This set of equations admits non-zero solutions only if $k$ satisfies
\begin{equation}\label{eq:quant1D}
  1-rr_0 \rme^{2\rmi kl_0} - rr_1 \rme^{2\rmi kl_1}+(r^2-t^2)r_0r_1 \rme^{2\rmi k(l_0+l_1)}=0 \, ,
\end{equation}
leading to the quantization condition of the quantum graph.

\begin{figure}[t]
  \centering
  \includegraphics[width=0.49\textwidth]{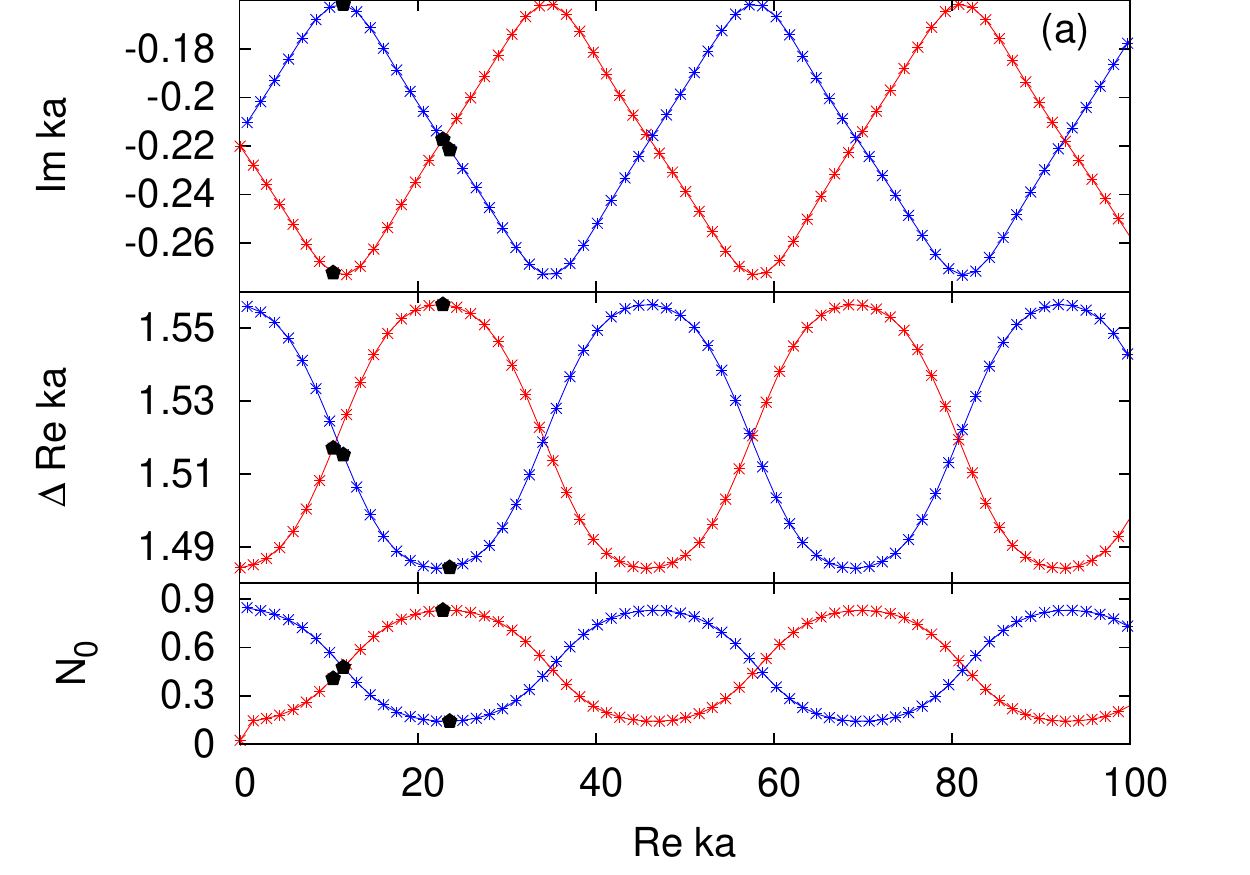}
  \includegraphics[width=0.49\textwidth]{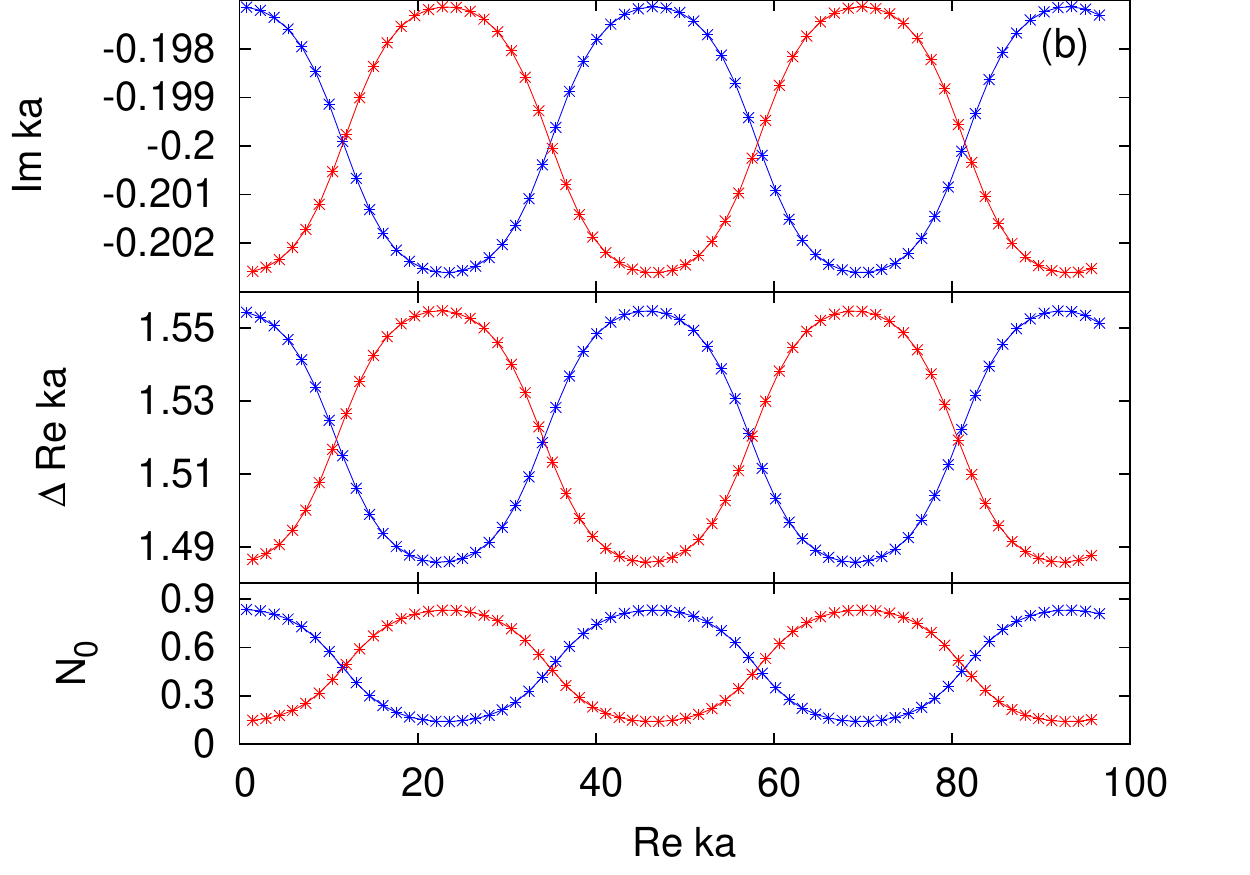}
\caption{\label{fig:k3-k17}
Panel (a) shows the resonance chains of the quantum graph for a generic $k$-independent value ${\cal F}=-0.63+\rmi0.79$, corresponding to the typical scenario where the oscillations of $\mathrm{Im}\,k_n$ (top) are out of phase from the oscillations of $\Delta \mathrm{Re}\,k_n$ (middle) and ${\cal N}_0$ (bottom). The black dots mark the resonances, for which the Husimi distributions are presented in \fref{fig:HusiFig}. Panel (b) illustrates the special case ${\cal F}=-1$, where these quantities all oscillate in phase.}
\end{figure}

We now observe that this condition approximately matches up with the cycle expansion expression \eref{eq:quantCE}.
In order to translate between both models we write $\rme^{\rmi k(L_{01}-L_0-L_1)}={\cal F}(k) $ and set $r=1$ (the motivation for this choice is discussed at the end of this section). One then can express all the parameters of the quantum graph in terms of the parameters of the cycle expansion. In particular, we find $l_0=L_0/2$, $l_1=L_1/2$, so that orbit 0 corresponds to running up and down edge 0, orbit 1 corresponds to running up and down edge 1, and orbit 01 approximately corresponds to running up and down the whole graph. Furthermore,
\begin{equation}\label{eq:mapping}
r_0=-\frac{1}{\sqrt{\Lambda_0}}, \quad
r_1=\frac{1}{\sqrt{\Lambda_1}}, \quad t=\left(\frac{\Lambda_{01}}{\Lambda_0\Lambda_1}\right)^{1/4}{\cal F}^{1/2}(k).
\end{equation}
Let us now exploit the implications of the correspondence between the cycle expansion and the quantum graph. Including the full $k$-dependence of all parameters, both produce the same resonance spectrum (see \fref{fig:figDisks}). Note that from $L_{01}\approx L_0+L_1$ the function ${\cal F}(k)$ is only slowly varying. Thus it is justified to locally neglect the $k$ dependence and approximate ${\cal F}(k)\approx {\rm const}$. This approximation would also be the natural choice for the quantum graph model as it corresponds to a $k$-independent scattering matrix. Setting ${\cal F}$ to a generic complex value with $|{\cal F}(k)|\approx 1$ [illustrated for ${\cal F}= -0.63+\rmi0.79$ in the top panel of \fref{fig:k3-k17}(a)], we can indeed reproduce the typical crossing of resonance chains displayed both by the cycle expansion (see \fref{fig:figDisks}) as well as by the full numerical results of the 3-disk system (see \fref{fig:Res_chains}, in the range $0<\mathrm{Re}\,ka <100$). Furthermore, we also reproduce the oscillations of $\Delta\,\mathrm{Re}\,k_na$, with maxima and minima coinciding with the crossings of the chains [middle panel of \fref{fig:k3-k17}(a)]. This behavior is observed for almost all choices of ${\cal F}$. Only for $|\mathrm{Im}\,{\cal F}| \ll |\mathrm{Re}\,{\cal F}|$ we observe a deviation from the generic behavior as shown in \fref{fig:k3-k17}(b) for ${\cal F}=-1$, where crossings of the chains are now enforced at the crossings of $\Delta \mathrm{Re}\,k_na$. Moreover, the amplitude of the oscillations in the resonance chains is then much reduced, while the amplitude of oscillations in $\Delta \mathrm{Re}\,ka$ remains the same (note the different scales in panel (a) and (b)). This effect can also be found in the exact numerical results (and in the cycle expansion) in the region $\mathrm{Re}\,ka\sim 65$, which corresponds to the exceptional value ${\cal F}(65) \approx -1$. The chains of the exact quantum resonances have an extra crossing in this region and smaller oscillation amplitudes while the oscillations in $\Delta \mathrm{Re}\,ka$ are not affected.

Another aspect which sets the quantum graph model apart from the cycle expansion is that it contains information on the resonance wave functions. This information is extracted by determining the amplitudes $(a_0^+,a_0^-,a_1^+,a_1^-)$ from \eref{eq:coefWave}, with $k=k_n$ set to a solution of the quantization condition \eref{eq:quant1D}. We normalize the coefficients such that $\int_{-l_0}^{l_1}|\psi(x)|^2\rmd x=1$, and then define ${\cal N}_0=\int_{-l_0}^{0}|\psi(x)|^2\rmd x$,
${\cal N}_1=\int_0^{l_1}|\psi(x)|^2\rmd x=1-{\cal N}_0$ as the probabilities to be situated on edge 0 or edge 1, respectively.
The bottom panels of \fref{fig:k3-k17}(a) and (b) show ${\cal N}_0$ for the two resonance chains. In both cases, the oscillations of ${\cal N}_0$ remain in phase with the oscillations in $\Delta \mathrm{Re}\,k_n$. This agrees well with the behaviour of the similarly defined asymmetry parameter \eref{eq:asym} in the 3-disk system, shown in \fref{fig:Asym}.

\begin{figure}
\centering
  \includegraphics[width=0.98\textwidth]{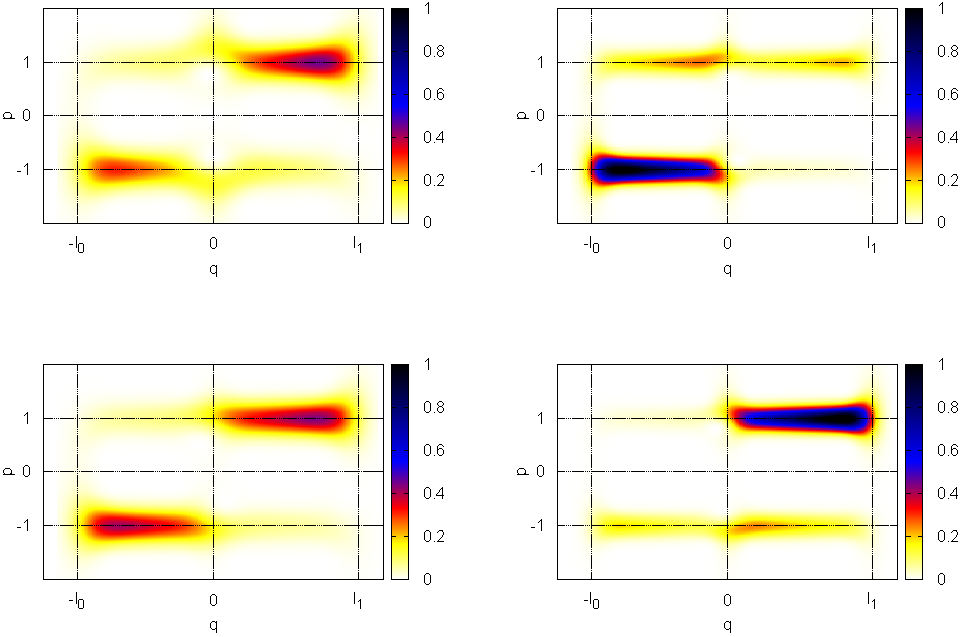}
\caption{\label{fig:HusiFig}
Husimi functions for the 8th (left) and 16th (right) resonance of the two resonance chains (top and bottom), for the same parameters as in \fref{fig:k3-k17}(a) (where the resonances are marked by black dots). The lengths $l_0$ and $l_1$ as well as the position of the connecting node are indicated by vertical lines. The horizontal lines mark the classical values of $p = \pm 1$.}
\end{figure}

\begin{figure}
  \centering
  \includegraphics[width=0.8\textwidth]{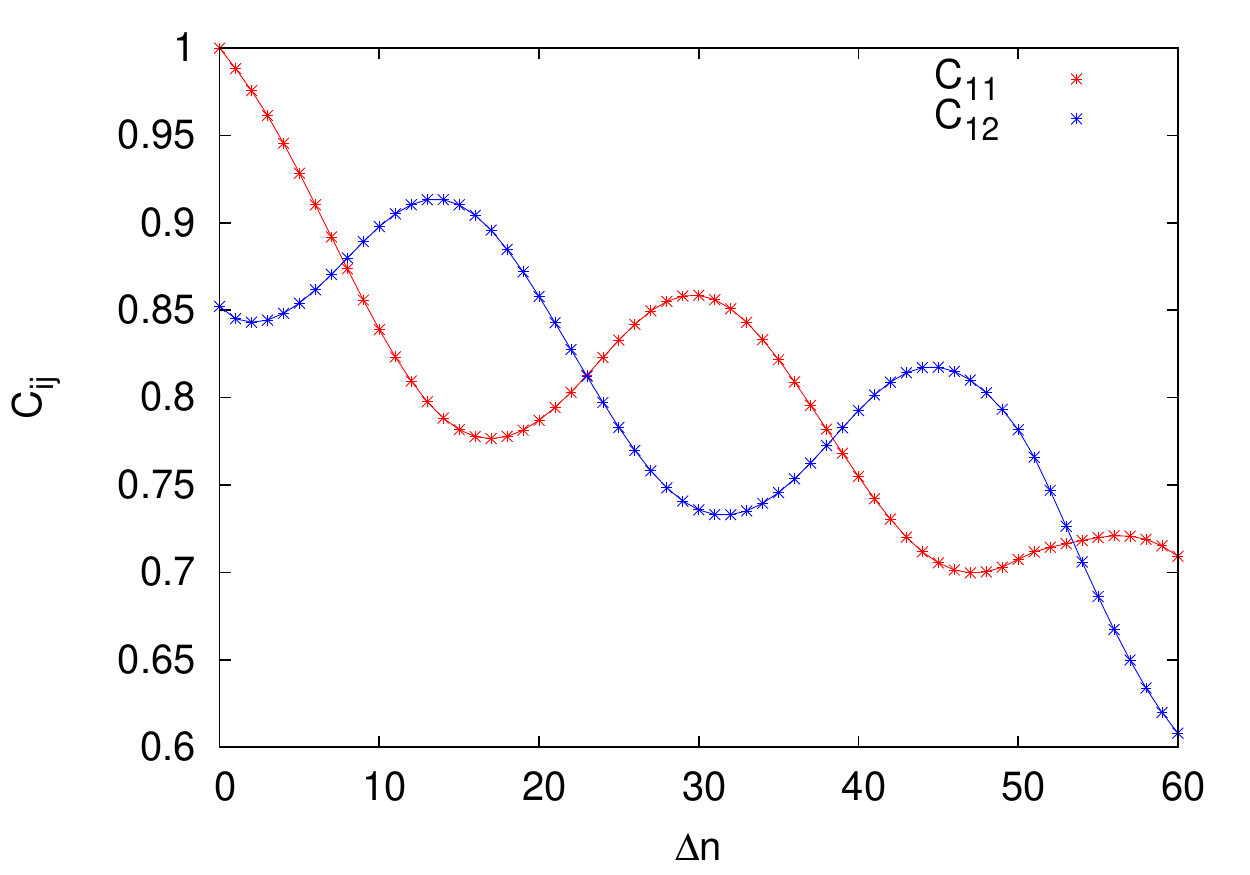}
\caption{\label{fig:corrQG}
Correlations $C_{11}$ and $C_{12}$ of the Husimi functions in the quantum graph model, with the same parameters as in \fref{fig:k3-k17}(a) ($\mathrm{max}-\mathrm{min}=30$).}
\end{figure}

This qualitative agreement also extends to the behavior of the Husimi functions. The Husimi functions are now defined by a convolution of the wave function with a conventional coherent state,
\[
 H_n^B(q,p)\propto|\langle c_{(q, p),\mathrm{Re}\,k_n} \mid \psi_n \rangle|^2\, ,
\]
where
\[
 c_{(q, p),k}(x)\propto \rme^{-\frac{1}{2}k(x-q)^2}\rme^{\rmi k p(x-q)}\, .
\]
The Husimi functions are normalized to $\int \rmd q \rmd p |H_n^B(q,p)|=1$, while the definition of the correlation function coincides with \eref{eq:Correlfunc}. For illustration, some Husimi functions of resonances in the quantum graph are shown in \fref{fig:HusiFig}. The chosen resonances (marked by black dots in \fref{fig:k3-k17}) illustrate once more the oscillating localization behavior on the two edges. Furthermore, the auto-correlations and cross-correlations (see \fref{fig:corrQG}) display the expected complementary oscillations, analogously to \fref{fig:Correl}.

To conclude this section we briefly discuss two general features of the quantum graph model. First, we note that the mapping of parameters \eref{eq:mapping} is not unique; it relies on our choice to set $r=1$. With this choice, the reflection coefficients $r_0$ and $r_1$ only depend on the properties of orbit 0 and orbit 1, respectively, which appears natural as one could consider the limit of a system in which these orbits are independent of each other (this corresponds to a putative system with $t=0$, where the two edges are not coupled). This particular choice of $r=1$ also implies, that the scattering matrix between the two edges is not unitary. Recall however, that we only find exact correspondence with the cycle expansion when we take the $k$-dependence of the matching coefficients into account. Thus our $k$-independent scattering matrix corresponds to an approximation of the scattering matrix which has already been meromorphically continued away from the real axis and it is therefore natural that it is nonunitary. In principle, however, the mapping is not unique, since the parameters in the quantum graph can be scaled as $(r,t,r_0,r_1)\to (\alpha r,\alpha t,\alpha^{-1} r_0, \alpha^{-1}r_1)$, for some (possibly complex) constant $\alpha$, without changing the quantization condition \eref{eq:quant1D}. The wave amplitudes then scale according to $(a_0^+,a_0^-,a_1^+,a_1^-)\to (a_0^+,\alpha a_0^-,\alpha a_1^+,a_1^-)$ (modulo overall normalization, $\int_{-l_0}^{l_1}|\psi(x)|^2\rmd x=1$).
 Therefore, this freedom only affects the relative weight of the two counter-propagating components on each edge, but does not mix the amplitudes of different edges. Indeed, we verified that the probabilities ${\cal N}_0$ and the correlations of the Husimi function become insensitive to this freedom in the translation once $\mathrm{Re}\,k$ is large, as is the case throughout the region of interest.

\begin{figure}[t]
  \centering
  \includegraphics[width=0.8\textwidth]{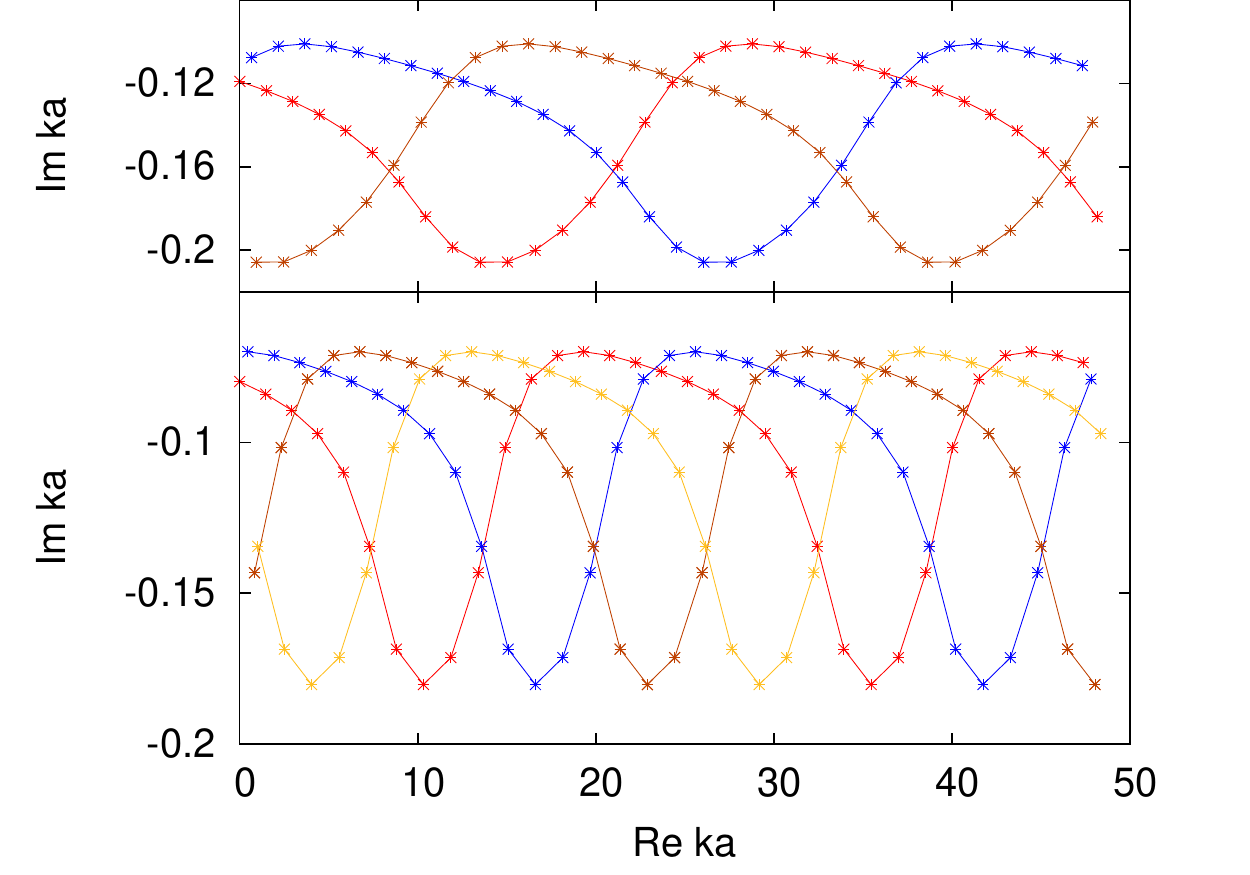}
\caption{\label{fig:deltaL}
Resonance chains of the quantum graph with the same stabilities as in \fref{fig:figDisks}, $l_0=2$ and $k$-independent ${\cal F}= -0.63+\rmi0.79$ and different $l_1$. For $l_1=4.25$ (top) and $l_1=6.5$ (bottom), corresponding to $n=2$ and $n=3$, leading to 3 and 4 chains respectively.}
\end{figure}

Secondly, it is worth pointing out that the fact that we describe two interacting resonance chains is not directly related with the existence of two edges in the quantum graph. The actual origin of the two chains is the fact that $l_0\approx l_1$. In general, we observe that $n+1$ chains appear when $l_1\approx n l_0$, indicating a surprising richness of the resonances in this simple quantum graph (see \fref{fig:deltaL}). By extension, we would expect to find a similar richness of resonance chains in quantum chaotic systems in which the leading order of the cycle expansion is given by the interaction of two short orbits of approximately commensurable length. Possible candidates fulfilling this requirement are non-symmetric 3-disk systems or Schottky surfaces \cite{arXbor13}.

\section{Conclusions}
\label{sec:conclusions}

In this article we linked the properties of interacting resonance chains in an autonomous open quantum-chaotic system, the 3-disk system, to underlying phase space structures consisting of trapped sets and periodic orbits. This allowed us to show that modulations and correlations of various spectral and spatial properties of the states along the chains arise from the interplay between different periodic orbits. Furthermore, we found first indications that the successive merging of resonance chains in the semiclassical limit, is connected with the increasing phase space resolution of the trapped set, thereby connecting the ubiquitous phenomenon of resonance chains with a fundamental physical mechanism that is also at play in the fractal Weyl law.

At the wave numbers accessed in this work, only the first order of the trapped set becomes clearly resolved. It is therefore desirable for future works to delve deeper into the semiclassical limit where higher orders can be resolved. Only in this regime one encounters the full range of questions related to to spectral gaps of semiclassical \cite{ika88,gas89a,non09,nau13,nau05,pet10} and random-matrix \cite{fyo97,som99,sch00} origin (recently observed in experiments \cite{bar13b}), mode non-orthogonality \cite{cha98,sch00,fyo02,kea08,pol09b,fyo12,sav13} (determining the line widths of microcavity lasers \cite{pat00,sch09b,cho12}), and local spectral statistics (presently far better understood in systems that obey random-matrix theory \cite{kuh08,pol09a,pol12}, which however cannot explain the fractal Weyl law as it assumes a vanishing Ehrenfest time \cite{sch05}). Similarly, we restricted the analysis of the cycle expansion to the second order in stability and the graph to two edges, which limits our phenomenological analogy with quantum graphs to periodic interactions of chains. Finally, it is also desirable to confirm the general picture developed here for other systems, including systems with two fundamental orbits of different near-commensurability for which the simple quantum-graph model already predicts interactions of a larger number of chains.

\ack
We thank S.~Nonnenmacher, A.~B\"{a}cker and B.~Eckhardt for helpful discussions. This work was supported by the Forschergruppe 760 `Scattering systems with complex dynamics'. TW acknowledges financial support of the `German National Academic Foundation', UK the support by the CNRS-INP via the program PEPS-PTI, and CP and HS the support by EPSRC via grant EP/J019585/1 ``Orbit-Based Methods for Multielectron Systems in Strong Fields''.

\section*{References}
\label{sec:references}

\end{document}